\documentclass[twocolumn,pdflatex,sn-nature]{sn-jnl}
 
\usepackage{geometry}
\usepackage{setspace}
\usepackage{graphicx}%
\usepackage{multirow}%
\usepackage{amsmath,amssymb,amsfonts}%
\usepackage{amsthm}%
\usepackage{mathrsfs}%
\usepackage[title]{appendix}%
\usepackage{xcolor}%
\usepackage{colortbl}
\usepackage{textcomp}%
\usepackage{manyfoot}%
\usepackage{booktabs}%
\usepackage{algorithm}%
\usepackage{algorithmicx}%
\usepackage{algpseudocode}%
\usepackage{listings}%
\usepackage[UTF8]{ctex}
\renewcommand{\figurename}{Figure}
\renewcommand{\abstractname}{Abstract}
\theoremstyle{thmstyleone}%
\theoremstyle{thmstyletwo}%

\theoremstyle{thmstylethree}%

\begin{document}

\title[Article Title]{Bridging scalp and intracranial EEG in BCI via pretrained neural representations and geometric constraint embedding}


\author[1,2]{\fnm{Yihang} \sur{Dong}}
\author[1]{\fnm{Changhong} \sur{Jing}}
\author*[1,2]{\fnm{Shuqiang} \sur{Wang}}\email{sq.wang@siat.ac.cn}

\affil[1]{\orgdiv{Shenzhen Institutes of Advanced Technology}, \orgname{Chinese Academy of Sciences}, \orgaddress{\city{Shenzhen}, \country{China}}}

\affil[2]{\orgname{University of Chinese Academy of Sciences}, \orgaddress{\city{Beijing}, \country{China}}}

\abstract{Electroencephalography (EEG) has become one of the key modalities underpinning brain–computer interfaces (BCIs) due to its high temporal resolution, rapid responsiveness, non-invasiveness, low cost, and portability. However, EEG signals are substantially inferior to intracranial EEG (iEEG) in signal-to-noise ratio and local spatial resolution, whereas iEEG suffers from extremely limited clinical accessibility owing to its invasive nature, hindering widespread application. To address this challenge, this study proposes a unified data- and prior knowledge–driven framework for EEG–iEEG representational enhancement. Guided by the principle that “geometric structure dictates function”, the framework maps static cortical anatomy onto dynamic constraints governing neural signal propagation and integrates general-purpose neural representations extracted by a pre-trained large EEG model to explicitly model signal transmission through the brain. Enhanced EEG signals are then synthesized via a multidimensional representation diffusion process. Numerous experimental results demonstrate that the generated enhanced EEG signals effectively recover the neural activity patterns lost during propagation through the brain. This finding indicates that the performance ceiling of BCIs is constrained not only by acquisition hardware but also by the depth to which the generative model resolves the mechanisms of neural signal propagation. Collectively, the proposed framework provides a viable pathway toward acquiring high-fidelity neural signals at low cost.}

\maketitle

\begin{spacing}{0.99999}
In recent years, BCI technologies have demonstrated substantial application potential in neurorehabilitation, human–machine interaction, and defense domains, a potential that hinges critically on accurate decoding of neural activity~\cite{chaudhary2016brain, schwemmer2018meeting, d2025towards, lee2025brain, merk2025invasive}. Multiple neuroimaging modalities including functional magnetic resonance imaging (fMRI), functional ultrasound (fUS), and functional near-infrared spectroscopy (fNIRS) have been employed for BCI decoding~\cite{rybavr2025simultaneous, griggs2024decoding}; however, these modalities generally suffer from insufficient temporal resolution, slow response latency, high instrumentation costs, and stringent acquisition environments, rendering them ill-suited for real-time, dynamic neural decoding. In contrast, EEG has emerged as the preferred signal source for non-invasive BCI owing to its millisecond-scale temporal resolution, rapid responsiveness, non-invasiveness, low cost, and high portability~\cite{ding2025eeg, huang2026robust, daly2023neural}. Intracranial EEG (iEEG) encompassing electrocorticography (ECoG), multielectrode arrays, and stereoelectroencephalography (sEEG) is regarded as the cornerstone modality for invasive BCI due to its direct recording of cortical electrical activity and exceptional signal-to-noise ratio~\cite{stolk2018integrated, dipalo2018plasmonic}. Nevertheless, EEG signals are inherently limited by volume conduction and skull attenuation, and are highly susceptible to artifacts from multiple sources. Conversely, iEEG requires craniotomy, resulting in extremely limited clinical accessibility~\cite{parvizi2018human}. These constraints severely impede the advancement of non-invasive and invasive BCI in terms of signal fidelity and widespread applicability, respectively. Consequently, enhancing non-invasive EEG signals to approximate the neural activity representational capacity of iEEG, thereby yielding enhanced EEG signals, constitutes a pivotal step toward realizing high-performance, non-invasive, low-cost, and portable BCI systems.

However, the key of EEG signal enhancement does not lie in investigating the neural sources or their activation states, but rather in learning the neural activity representations lost during signal propagation. The former, which aims to localize the origin of neural electrical activity, generally falls under EEG source imaging and fundamentally addresses the question “where is the neural activity generated?”~\cite{michel2004eeg, kaiboriboon2012eeg}. In contrast, this study aims to generate high-fidelity enhanced EEG signals to improve performance on downstream BCI tasks. Recently, Bahman Abdi-Sargezeh and colleagues attempted to map EEG directly to iEEG using deep neural networks in an end-to-end manner~\cite{abdi2021higher, abdi2023mapping,abdi2025eeg}. Although these studies preliminarily demonstrate the potential of data-driven approaches in modeling complex nonlinear mappings, their task formulation is inherently limited to signal-level fitting and cannot achieve generalizable EEG signal enhancement. On the other hand, Miao Cao and colleagues proposed reconstructing iEEG from magnetoencephalography (MEG) signals using an LCMV beamformer~\cite{cao2022virtual}. However, this approach remains constrained by linear assumptions and simplified brain models, yielding reconstructions insufficient for high-precision BCI applications. More critically, MEG systems are prohibitively expensive, require stringent magnetically shielded environments, and are impractical for bedside or long-term monitoring scenarios. In comparison, EEG offers superior portability, broad accessibility, and low cost, making it a more viable modality for real-world BCI deployment~\cite{santhanam2006high}.

Unlocking this modality's potential requires overcoming limitations in modeling physical constraints. The brain exhibits intrinsic structure-function coupling, with cortical geometry as the scaffold shaping neural activity and encoding biophysical boundary conditions governing signal generation, propagation, and integration~\cite{wang2024human, pun2024measuring}. Geometric modal decomposition maps static anatomy into dynamic functional priors, revealing embedded spectral properties and establishing a principled pathway for structure-guided signal enhancement~\cite{jiang2025gene,zhang2025dynamic}. Incorporating geometric priors compensates for information loss in non-invasive recordings, shifting reconstruction from data-driven toward a structure–function coupled paradigm and laying a theoretical foundation for next-generation high-fidelity EEG enhancement.

\begin{figure*}[htbp]
    \centering
    \includegraphics[width=\linewidth]{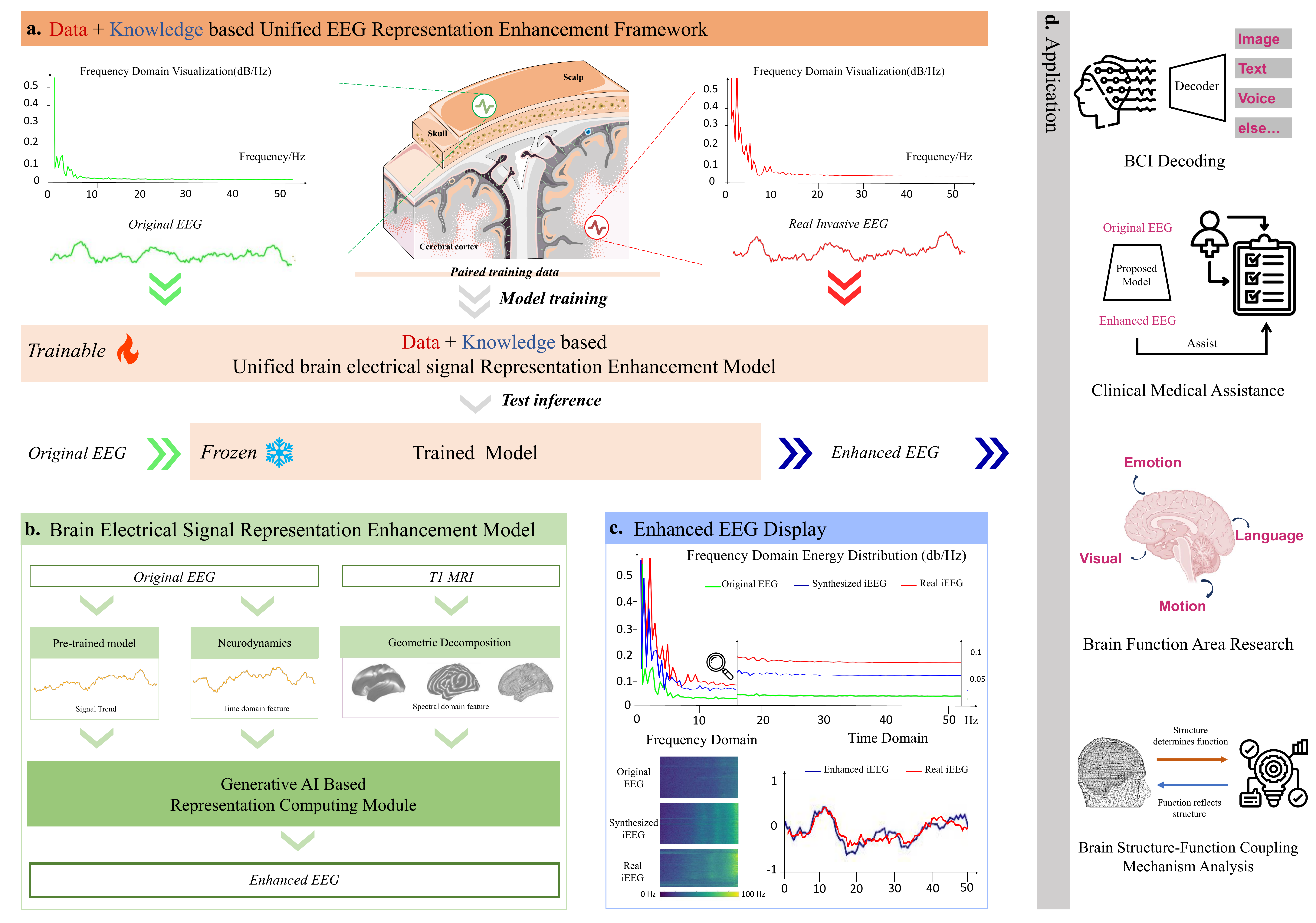}
    \caption{Schematic of the proposed framework. a, Data- and knowledge-based unified EEG representation enhancement framework. The model is trained with scalp EEG as input and iEEG as the target; during inference, enhanced EEG signals are generated from scalp EEG alone. b, Detailed architecture of the enhancement model, integrating a data-driven module, a neurodynamical–geometric constraint module, and a generative AI–based representation module. Individualized T1-weighted MRI can be incorporated alongside EEG to further improve performance. c, Multidimensional visualization of enhanced EEG signals, including spectral power distribution, spatial maps of band-specific activity, and temporal waveform morphology, collectively demonstrating recovery of neurophysiologically relevant features. d, Applications of the proposed framework, encompassing improved BCI decoding accuracy, enhanced BCI fairness , and support for brain functional mechanism research.}
    \label{fig:fig1}
\end{figure*}

To address these challenges, we propose a unified EEG–iEEG representation enhancement framework that integrates data-driven learning with neurodynamical priors, as shown in Figure~\ref{fig:fig1}. The framework comprises four core components. First, a pretrained large EEG model is employed to extract general-purpose EEG representations, thereby strengthening the modeling capacity for neural activity patterns~\cite{jiang2024large, cui2024neuro}. Second, explicit biophysical constraints are incorporated through joint modeling of neural mass models and neural field equations, ensuring that the enhanced signals adhere to the electrophysiological dynamics governing brain activity~\cite{scheeringa2011neuronal, kiebel2006dynamic}. Third, to account for the complex geometric morphology of the brain, the proposed method processes T1-weighted MRI data via geometric mode decomposition to compute structurally informed priors across frequency bands, which model the attenuation of neural signals during propagation through skull and brain tissues and thereby enable recovery of lost neural representations during enhancement~\cite{pang2023geometric, yu2018geometric}. Fourth, a multidimensional EEG feature encoder fuses the above three elements—representations, dynamical constraints, and structural priors—to generate features that precisely capture the coupling between long-range neural synchrony and local high-frequency activity; a diffusion model then guides the generation of enhanced EEG signals to improve their intrinsic fidelity to underlying neural activity patterns. 


Systematic evaluation across 14 diverse EEG datasets spanning multiple task types including emotion recognition shows that the proposed EEG–iEEG unified representational enhancement framework significantly improves the fidelity with which scalp EEG captures intracranial neural activity, without requiring invasive recordings or hardware upgrades. This work thus enables low-cost acquisition of high-fidelity enhanced EEG signals and provides a new technical pathway to overcome the performance bottlenecks of current non-invasive BCI.

\end{spacing}

\section*{Results}\label{sec2}

\subsection*{Consistency evaluation between synthesized iEEG and ground-truth iEEG}\label{subsec2}
This study aims to enhance EEG signals using the proposed framework to generate high-fidelity synthesized iEEG and evaluate its consistency with real iEEG. Given that neural activity representations inherently couple temporal dynamics, spectral energy distribution, and spatial propagation characteristics, similarity along a single dimension is insufficient to fully validate the physiological plausibility of the generated signals. Therefore, this set of experiments is conducted across multiple complementary dimensions: spectral, temporal, time–frequency joint structure, and inter-channel correlation. Specifically, in the spectral domain, power spectral density similarity (PSD Sim) and spectral feature similarity (Freq Sim) are used to assess consistency in energy distribution and rhythmic structure~\cite{youngworth2005overview}; in the temporal domain, Pearson correlation coefficient (PCC) and cosine similarity (Cos Sim) quantify the alignment of waveform morphology and dynamic evolution trends~\cite{benesty2009pearson, xia2015learning}; additional metrics include phase consistency to capture time–frequency coupling properties and Frobenius norm similarity of inter-channel correlation matrices (CorrMat Sim) to reflect connectivity fidelity at the channel level~\cite{thatcher2005eeg, schindler2007assessing}. The results demonstrate that the proposed framework effectively enhances EEG signals and recovers the neural activity representations lost during signal transmission through the brain’s anatomical structure.

\begin{figure*}[htbp]
    \centering
    \includegraphics[scale=0.31]{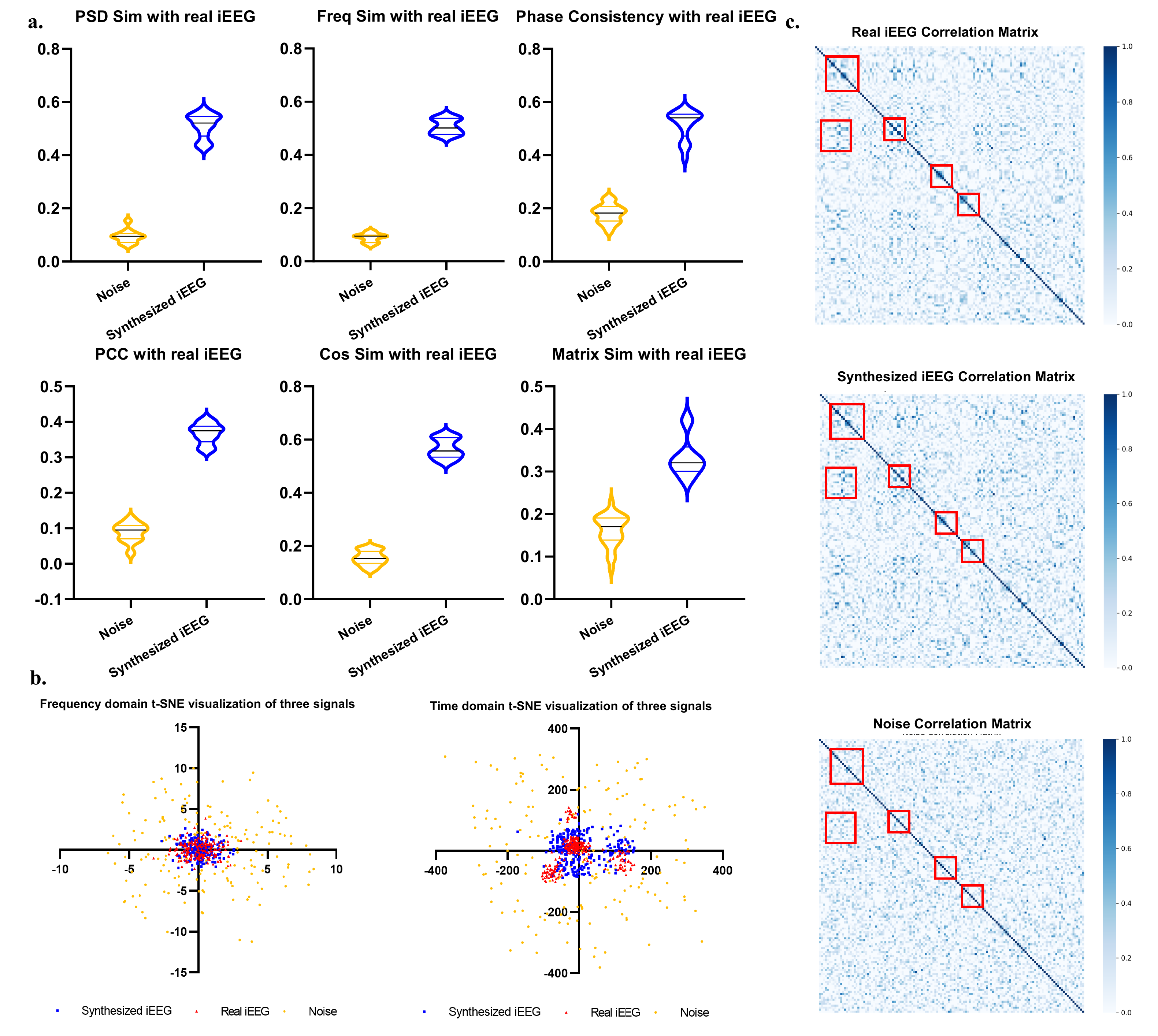}
    \caption{Multidimensional validation of consistency between synthesized iEEG and real iEEG. a, Across six complementary metrics—spectral (PSD Sim, Freq Sim), temporal (PCC, Cos Sim), and time–frequency coupling (Phase Consistency, CorrMat Sim)—synthesized iEEG generated from EEG (blue) significantly outperforms noise-driven controls (yellow), demonstrating effective recovery of high-dimensional neural dynamics lost to skull attenuation and volume conduction. b, t-SNE embeddings in the time–frequency domain show tight clustering of synthesized iEEG with real iEEG (red), while noise-driven signals (yellow) are distinctly separated, confirming acquisition of a physiologically plausible joint time–frequency representation. c, Heatmaps of full-channel correlation matrices reveal that synthesized iEEG recapitulates the local clustered and long-range synchronized connectivity patterns observed in real iEEG within five key functional networks, whereas noise-driven reconstructions lack such neurophysiologically specific topological structure.}
    \label{fig:fig2}
\end{figure*}

As shown in Figure~\ref{fig:fig2}a, synthesized iEEG guided by real EEG (blue) significantly outperforms the noise-guided control signal (yellow) across all six evaluation metrics. In the spectral domain, PSD Sim and Freq Sim reach 0.51 and 0.50, respectively—both consistently exceeding the physiologically meaningful threshold of 0.5 and improving by more than 0.4 over the noise-guided baseline. This result directly confirms that the framework effectively compensates for high-frequency energy loss caused by skull attenuation and reconstructs attenuated rhythmic spectral structures. Further analysis of temporal characteristics reveals a PCC of 0.38 and a Cos Sim of 0.57, indicating strong alignment between synthesized iEEG and real iEEG in waveform dynamics and instantaneous amplitude distribution. In the additional dimensions, Phase Consistency reaches 0.52, further validating the physiological plausibility of synthesized iEEG in time–frequency coupling; its phase synchrony not only markedly surpasses that of the noise-guided signal but also closely matches the neural oscillatory dynamics of real iEEG. Collectively, the coordinated improvement across multiple metrics demonstrates that the proposed framework successfully learns the high-dimensional dynamic representations lost during neural signal transmission through brain anatomy.

To assess the representational capacity of the reconstructed signals, Figure~\ref{fig:fig2}b presents t-SNE visualizations of three signal types~\cite{maaten2008visualizing}: real EEG–guided synthesized iEEG (blue), real iEEG (red), and noise-guided iEEG (yellow), labeled as Synthesized iEEG, Real iEEG, and Noise, respectively. In the time–frequency embedding space, synthesized iEEG clusters tightly with real iEEG, whereas the noise-guided signal is distinctly separated. This result demonstrates that the proposed framework effectively learns a general time–frequency representation distribution closely aligned with real iEEG and substantially enhances the original EEG signal, markedly outperforming the unstructured, noise-guided generation approach.

Finally, Figure~\ref{fig:fig2}c displays heatmaps of full-channel correlation matrices for the three signal types, with focused comparison in five regions exhibiting prominent functional connectivity. The synthesized iEEG generated under real EEG guidance successfully reproduces the local clustered connectivity and long-range cross-regional synchronization patterns observed in real iEEG. In contrast, the noise-guided signal lacks such structure, exhibiting diffuse correlations without discernible topological features. This finding provides strong evidence that the proposed enhancement framework faithfully reconstructs the inter-channel structural topology characteristic of iEEG, achieving high-fidelity emulation of cortical electrophysiological activity.

\subsection*{Performance gains of enhanced EEG signals in downstream tasks}\label{subsec3}
Experiments on twelve diverse BCI downstream task datasets—including motor imagery and cognitive detection—demonstrate that the proposed framework yields high-fidelity enhanced EEG signals at low cost and consistently improves task performance. Specifically, the framework was first applied to generate enhanced signals on twelve public EEG–BCI datasets, and the performance difference between original and enhanced signals was quantified under identical experimental conditions. Second, on the SEED benchmark dataset for emotion recognition, topographic maps were used to visualize and compare activation patterns in emotion-relevant brain regions before and after enhancement, thereby validating the physiological plausibility of the enhanced signals. Finally, SHAP (SHapley Additive exPlanations) analysis was employed to examine shifts in channel contribution rankings for the emotion classification task, providing further evidence that the enhanced signals deliver meaningful performance gains over raw EEG~\cite{lundberg2017unified}.

\begin{figure*}[htbp]
    \centering
    \includegraphics[width=\linewidth]{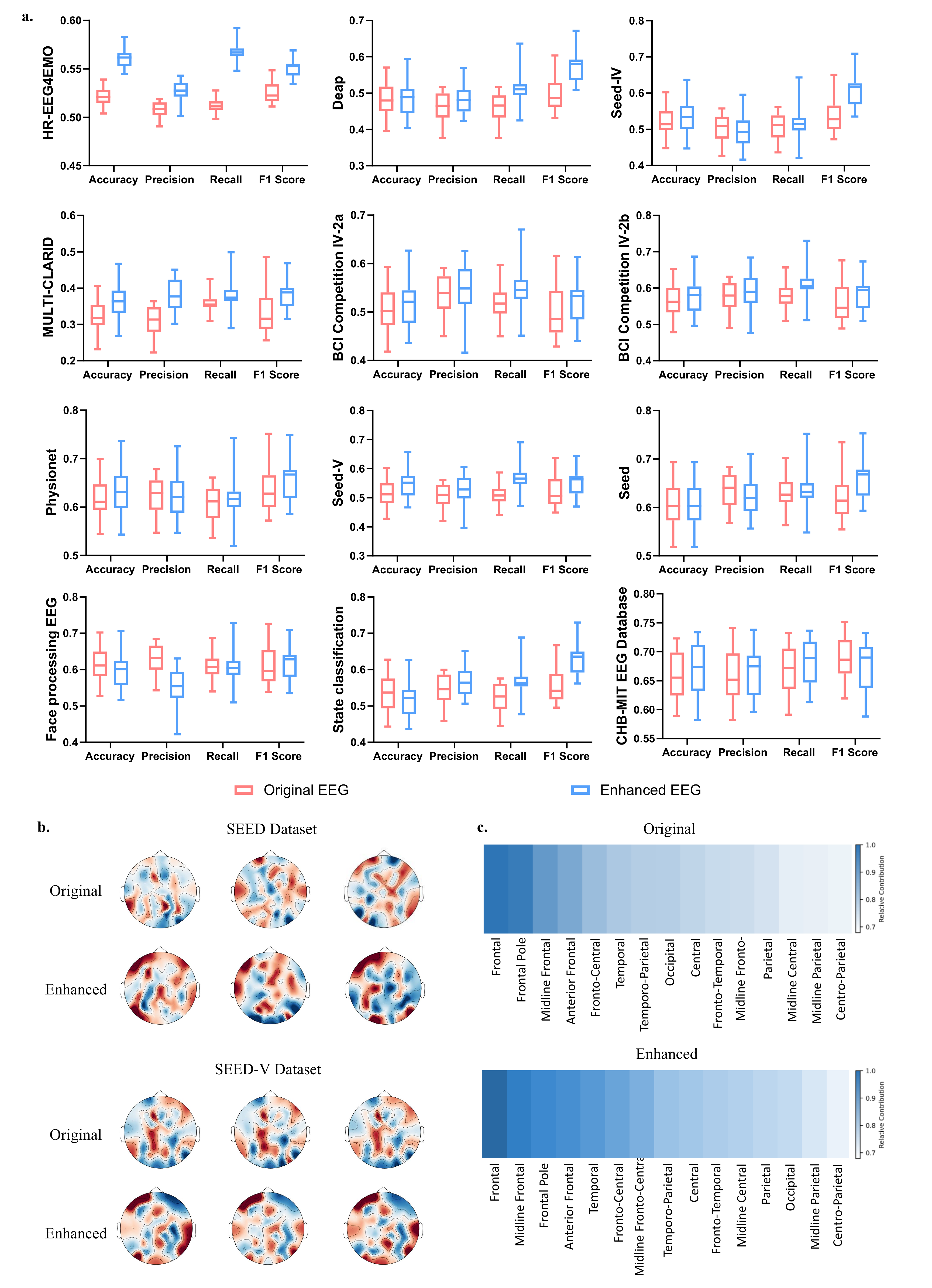}
    \caption{Performance gains and neurophysiological validity of enhanced EEG across diverse BCI tasks. a, Across 12 EEG datasets spanning motor imagery emotion and cognitive decoding enhanced EEG boosts classification F1 scores by 2.8\% to 8.6\% (mean 5.2\%) indicating recovery of high-frequency neural dynamics lost to tissue filtering. b, On SEED topographic maps show enhanced EEG yields focal high-intensity activation in dorsolateral prefrontal and bilateral temporal regions consistent with emotion circuitry whereas raw EEG shows diffuse weak activity. c, SHAP analysis reveals top channels in enhanced EEG such as Fp1 and Fz cluster in prefrontal limbic areas and their importance ranking correlates more strongly with emotion-related localization rising from 0.41 to 0.62.}
    \label{fig:fig3}
\end{figure*}

As shown in Figure~\ref{fig:fig3}a, enhanced EEG signals consistently improve downstream classification performance across all twelve multi-task EEG datasets, with gains ranging from 2.8\% to 8.6\%. Notably, the improvements are most pronounced in emotion and cognitive state decoding tasks: on the DEAP dataset for emotion classification, the F1 score increases by 8.6\%; in cognitive state classification, the gain reaches 8.5\%; and on the four-class SEED-IV emotion recognition task, performance improves by 8.1\%. Across all datasets, the proposed framework achieves an average performance gain of 5.2\%, demonstrating its ability to effectively recover high-frequency neural activity representations lost during signal propagation through complex brain tissues, including the skull, scalp, and cerebrospinal fluid, and thereby enhance the generalization capability of the generated EEG signals in downstream tasks.

To further validate physiological plausibility, a visualization analysis was conducted on the SEED emotion recognition dataset. Figure~\ref{fig:fig3}b displays topographic maps of raw and enhanced EEG under positive, neutral, and negative emotional states. Raw EEG exhibits diffuse, shallow-red activation across the entire scalp, indicating weak and spatially disorganized neural activity in emotion-relevant regions. In contrast, enhanced EEG reveals focal, high-intensity activation, manifested as concentrated deep-red clusters, in the dorsolateral prefrontal cortex and bilateral temporal lobes, areas well established in emotion processing~\cite{kragel2016decoding, lindquist2012brain, dalgleish2004emotional}. This pattern aligns closely with known neurophysiological mechanisms of emotion, confirming that the proposed framework enhances EEG signals in a manner consistent with underlying brain physiology while improving downstream task performance.

Finally, SHAP was employed to quantify the contribution of each EEG channel to the emotion classification task. As shown in Figure~\ref{fig:fig3}c, the top five channels with highest contribution in the enhanced signal, including Fp1 and Fz, are localized within the prefrontal limbic system, matching the canonical neuroanatomical substrates of emotion processing~\cite{kragel2016decoding, lindquist2012brain, dalgleish2004emotional}. In contrast, high-contribution channels in the raw signal are scattered across non-specific regions such as the parietal Pz. The Spearman correlation coefficient between channel rankings and the ground-truth emotion-related topography increases from 0.41 in raw EEG to 0.62 in enhanced EEG, directly demonstrating that the framework selectively learns task-relevant neural activity representations from physiologically meaningful regions.

\subsection*{Impact analysis of geometric constraints on iEEG signal reconstruction performance}\label{subsec4}
This study processes T1-weighted MRI using a geometric modal decomposition method to obtain a set of geometric features representing cortical curvature. This geometric feature set is incorporated into the proposed enhancement framework to explicitly model the complex geometric constraints of the cerebral cortex and the associated mechanisms of electrical signal transmission, thereby enabling the framework to capture high-dimensional dynamic neural information lost during propagation of neural signals through multi-scale brain structures~\cite{pang2023geometric}. This experimental series systematically evaluates the impact of the T1-MRI–based geometric modal decomposition on iEEG reconstruction performance through three core analyses: first, differential decomposition of the T1-MRI signal is performed and the resulting geometric features are visualized; second, under identical experimental conditions, each derived geometric feature set is fed into the proposed framework to generate synthesized iEEG, and its consistency with real iEEG is assessed using multiple quantitative metrics; third, for each geometric feature set comprising N features, an intra-set correlation matrix is computed to quantify feature redundancy, providing a qualitative basis for selecting the optimal value of N in geometric constraint modeling.

The incorporation of T1-weighted MRI structural information enables the proposed framework to preserve static anatomical specificity while explicitly modeling the modulatory effects of cortical curvature and topology on neural signal conduction, allowing the generated enhanced EEG signals to effectively approach the high-dimensional neural representational capacity of iEEG. This advancement further drives non-invasive BCI signal enhancement from empirical data fitting toward interpretable modeling grounded in structure–function coupling principles.

\begin{figure*}[htbp]
    \centering
    \includegraphics[scale=0.25]{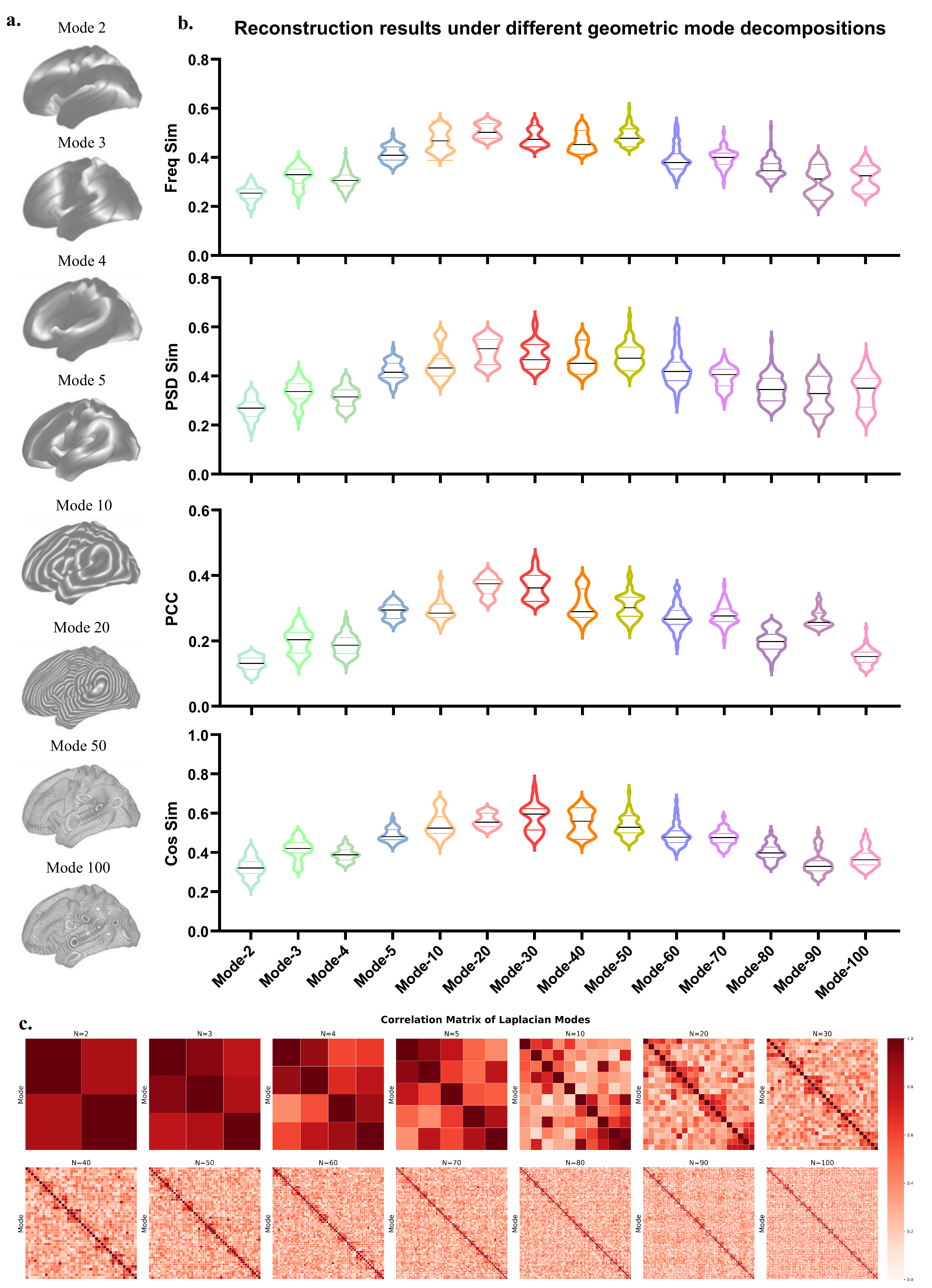}
    \caption{Impact of T1-MRI–derived geometric constraints on iEEG reconstruction fidelity. a, Eight geometric modal features extracted from T1 MRI show coarse large-scale patterns at low N (2–5) and progressively finer gyral-level detail as N increases to 20–100. b, Consistency between synthesized and real iEEG peaks at N = 20 (Freq Sim 0.50 PSD Sim 0.51 PCC 0.38 Cos Sim 0.57) then declines with further increases in N revealing an optimal precision threshold for geometric constraint modeling. c, Correlation matrix heatmaps illustrate feature redundancy at N $\le$ 10 strong diagonal dominance with moderate off-diagonal structure at N = 20–40 indicating balanced independence and physiological coherence and increasing noise and loss of structure at N $\geq$ 50 consistent with performance trends in b.}
    \label{fig:fig4}
\end{figure*}

Figure~\ref{fig:fig4}a displays the visualization of eight geometric feature sets obtained via cortical geometric mode decomposition. Features derived at low N values (N = 2–5) exhibit coarse-grained spatial patterns, primarily capturing large-scale cortical geometry. As N increases to 20–100, the features become progressively refined, resolving gyral- and sulcal-level anatomical details.

As shown in Figure~\ref{fig:fig4}b, the consistency between synthesized iEEG and real iEEG exhibits a non-monotonic trend with increasing N: performance initially rises, then fluctuates, and eventually declines, peaking at N = 20, where Freq Sim reaches 0.50, PSD Sim 0.51, PCC 0.38, and Cos Sim 0.57. This set of experiments demonstrates that an optimal precision threshold exists between cortical geometric constraint modeling and the overall framework; both insufficient and excessive numbers of geometric features degrade framework performance to some extent, thereby compromising the fidelity of the synthesized iEEG.

The correlation matrix heatmap in Figure~\ref{fig:fig4}c further reveals the interdependencies and structural properties among the cortical geometric modal features.. When N $\le$ 10, the matrices display deep-red, block-like regions of strong correlation, indicating high feature redundancy. At N = 20–40, the main diagonal becomes distinctly prominent while off-diagonal correlations remain moderate, reflecting a balance between structural independence and necessary physiological coupling. When N $\geq$ 50, the matrices exhibit highly disordered patterns with substantial noise and diminishing feature coherence, consistent with the quantitative performance decline observed in Figure 4b.

\subsection*{Contribution analysis of individual components in the enhancement framework}\label{subsec5}
The effectiveness of the proposed framework stems from its systematic integration of data-driven representations, neural dynamics constraints, and geometric structural priors. To dissect the mechanistic contributions of these components to EEG signal enhancement, ablation experiments were conducted focusing on three core elements: the Pretrained EEG Representation Encoder (PERE), the Neural Dynamics Constraint (NC), and the Geometric Constraint (GC). The experimental design comprises three parts. First, on a simultaneously recorded EEG iEEG dataset, eight ablation configurations were evaluated to quantify the individual and synergistic effects of each component on multi-dimensional reconstruction metrics, including spectral and temporal domains. Second, on the SEED emotion recognition dataset, topographic maps of enhanced EEG signals generated under each ablation setting were visualized to qualitatively assess whether the activation patterns align with known emotion-related cortical regions and to compare enhancement outcomes across configurations. Third, the impact of each ablation setup on downstream discriminative performance was evaluated on two classification tasks, MULTI-CLARID and SEED, to reveal the relative priority of each component in terms of neural information recovery and task adaptation. In this notation, “Full” denotes the complete enhancement framework incorporating PERE, NC, and GC, while all other settings are labeled as “w/o X” to indicate the removal of the corresponding component.

\begin{figure*}[htbp]
    \centering
    \includegraphics[scale=0.27]{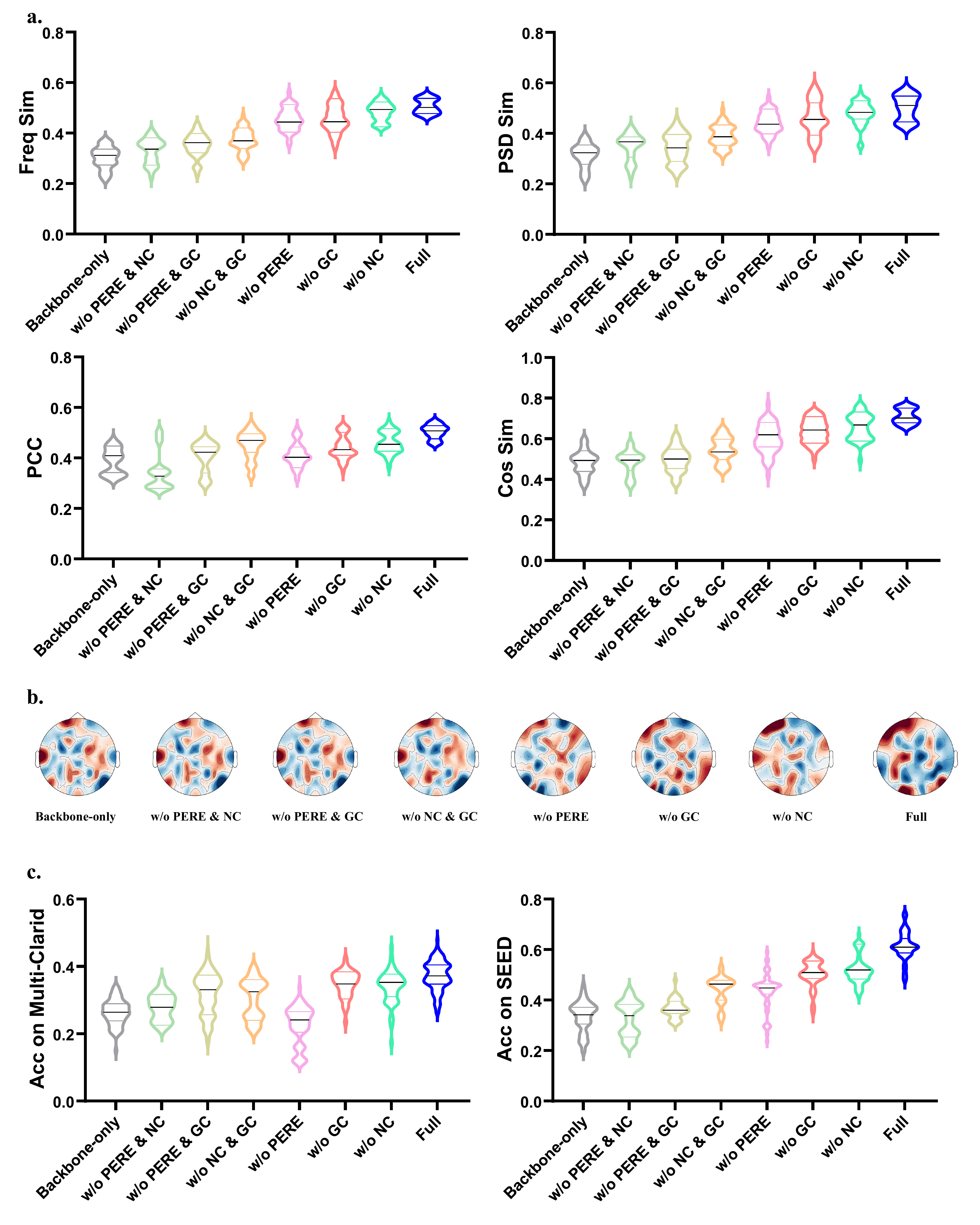}
    \caption{Ablation study of framework components. a, The full model (PERE + NC + GC) achieves highest fidelity across all metrics; removing PERE causes the largest drop while GC and NC provide complementary gains in spectral and phase consistency. b, Only the full model yields focal activation in emotion-relevant regions on SEED matching known neurophysiological patterns; ablated variants show diffuse activity. c, Downstream accuracy on MULTI-CLARID and SEED rises with component integration reaching 0.48 and 0.60 which are 84\% and 39\% above baseline confirming synergistic enhancement from data-driven representation neurodynamics and geometry.}
    \label{fig:fig5}
\end{figure*}

As shown in Figure~\ref{fig:fig5}a, the full framework achieves superior performance across all four metrics compared to every ablation variant. Specifically, Freq Sim and PSD Sim reach 0.50 and 0.51 under the full framework, representing an improvement of over 50\% relative to the backbone-only baseline; PCC and Cos Sim attain 0.38 and 0.57, respectively, also showing consistent gains. Ablation analysis reveals that removing PERE causes the largest performance drop, with reductions exceeding 0.08 in both Freq Sim and PCC, indicating that the general-purpose time–frequency representations provided by PERE form the foundational support of the framework. Removal of GC substantially degrades PSD Sim, highlighting the critical role of geometric constraint modeling in enhancing spectral fidelity. Although NC contributes relatively modestly when introduced alone, it consistently improves phase consistency in combination with other components, demonstrating its effectiveness in capturing the temporal evolution of neural electrical activity. These results confirm a nonlinear complementary relationship among the three components, where the absence of any single element leads to a marked decline in overall performance, thereby validating their irreplaceable roles within the framework.

Figure~\ref{fig:fig5}b displays topographic maps of enhanced EEG signals under eight ablation settings on the SEED dataset, where color intensity reflects channel activation strength. Darker red indicates higher activation and darker blue indicates lower activation. In the first four ablation conditions, activation patterns appear diffuse and lack spatial focus. In contrast, the latter four configurations show markedly more focal activation, particularly in the left temporal lobe and temporo-occipital junction~\cite{dalgleish2004emotional, olson2007enigmatic}. Notably, the topography generated by the full framework exhibits not only concentrated high-activation regions but also clear intensity gradients that closely match the spatial patterns of genuine emotion-evoked EEG. This qualitative observation aligns precisely with the quantitative results in Figure 5c, corroborating the effectiveness of each component in EEG signal enhancement.

Figure~\ref{fig:fig5}c presents performance changes on two downstream classification tasks, MULTI-CLARID and SEED. As PERE, GC, and NC are incrementally incorporated, model accuracy increases monotonically on both datasets. The full framework achieves 0.48 on MULTI-CLARID and 0.60 on SEED, corresponding to improvements of 84\% and 39\% over the backbone-only baseline, respectively. These findings demonstrate that the proposed framework establishes a multidimensional synergistic enhancement architecture by integrating general-purpose representations, neural dynamics constraints, and geometric structure modeling, thereby achieving both higher fidelity in enhanced EEG signals and robust performance gains across diverse downstream tasks.

\subsection*{Contribution analysis of EEG signals across frequency bands to enhancement outcomes}\label{subsec6}
To quantify the contribution of distinct neural oscillatory bands to EEG signal enhancement, this section systematically evaluates the reconstruction performance of the delta, theta, alpha, beta, and gamma frequency bands under the unified framework. Results reveal pronounced spectral specificity in the enhancement effect, with high-frequency components playing a dominant role in recovering localized neural activity and spatial organization.

\section*{Discussion}\label{sec3}
To address the performance ceiling of non-invasive brain–computer interfaces imposed by the measurement-centric paradigm of traditional neural engineering, this study reframes the generation of enhanced brain signals as a Bayesian inference process and proposes a unified EEG enhancement framework that jointly integrates cortical geometry and function. By incorporating T1-weighted MRI to capture the brain's geometric architecture and underlying physical laws, the framework establishes a structural foundation for representing functional brain patterns, enabling EEG enhancement without any modification to acquisition hardware and allowing enhanced EEG signals to approach the representational capacity of invasive recordings. This work marks a paradigm shift in BCI signal enhancement from black-box data fitting toward neural signal computation grounded in the principle that geometry shapes function.

The performance of the current framework remains constrained by several key factors. First, the scarcity of synchronized scalp EEG and iEEG paired recordings, combined with the limited scale and generalizability of existing pre-trained EEG models, jointly restrict the model’s capacity to learn complex neural propagation mechanisms and high-frequency dynamic features~\cite{jiang2024large}. Second, substantial heterogeneity across studies in experimental paradigms, acquisition protocols, and task designs introduces considerable electromyographic, ocular, and environmental noise during data collection, which often obscures underlying neural activity, severely degrades EEG signal-to-noise ratio, and consequently limits both the upper bound of enhancement efficacy and cross-task robustness~\cite{delorme2023eeg}. Third, the present framework incorporates neurodynamical constraints only at the mesoscale, focusing on mean-field modeling of neural population activity; microscale single-neuron dynamics and macroscale inter-regional network coupling have not yet been integrated due to challenges in task adaptability and computational complexity, thereby limiting its ability to fully capture multiscale neural dynamics. Collectively, these factors constrain further improvements in the fidelity of enhanced EEG signals and underscore the necessity of evolving from a “generic enhancement” paradigm toward one of “task- and subject-coordinated optimization.”

Future work could explore region-specific generative strategies targeting functionally defined brain areas, by integrating local anatomical priors with region-specific neurodynamical models to enable precise reconstruction and modulation of electrical activity in designated cortical regions~\cite{zhang2025harnessing}. This direction would shift EEG enhancement from “global signal reconstruction” toward “task-driven enhancement of specific neural activity,” thereby substantially improving the fidelity of downstream task modeling~\cite{olejarczyk2022region}. Moreover, such a strategy has the potential to significantly reduce reliance on high sensor density and large channel counts while maintaining high reconstruction fidelity, offering a critical technical foundation for next-generation non-invasive brain–computer interfaces that are lightweight, comfortable, and suitable for long-term wear.

\begin{figure*}[htbp]
    \centering
    \includegraphics[scale=0.40]{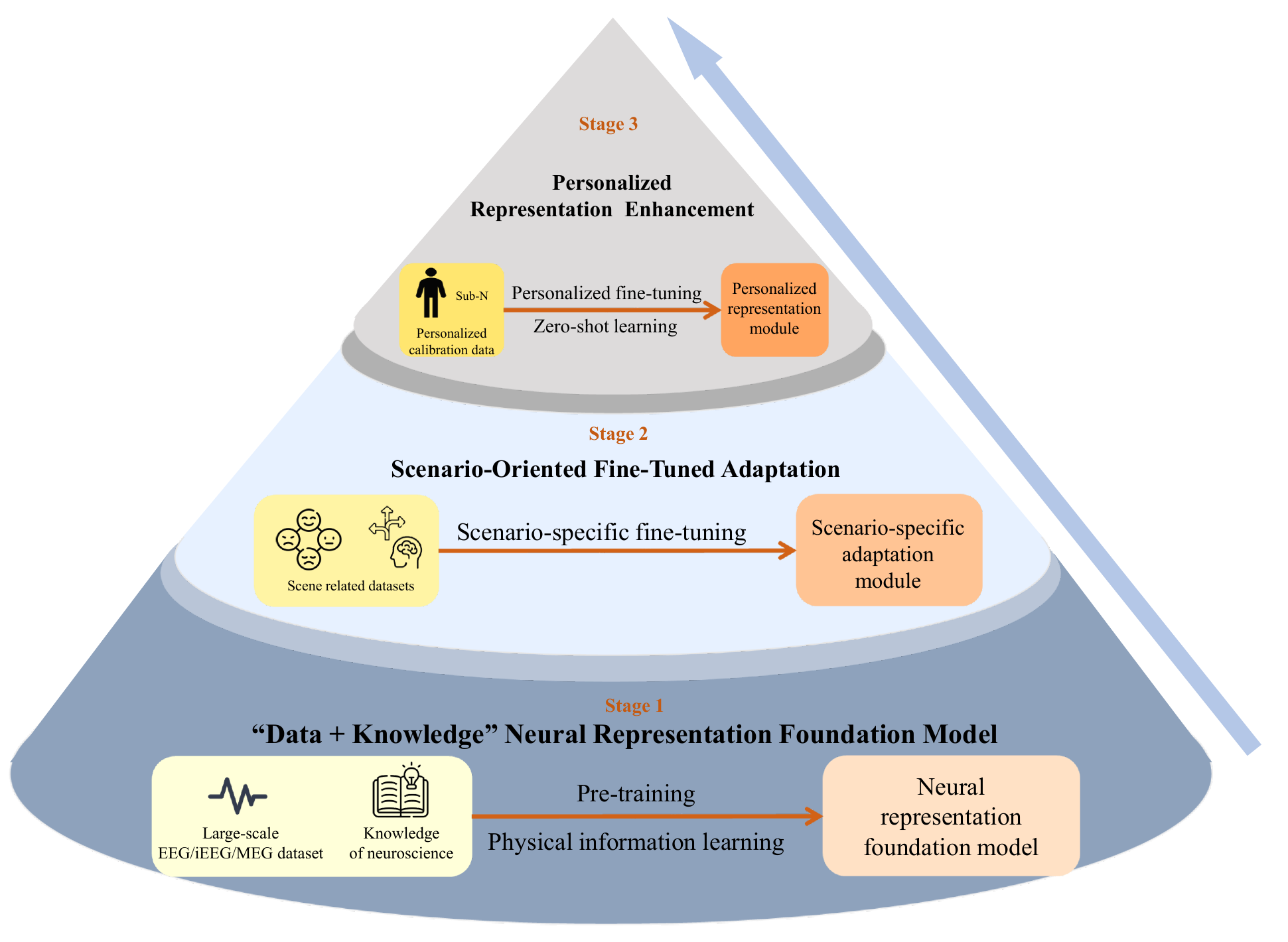}
    \caption{A three-stage paradigm for EEG enhancement. Stage 1 builds a neural representation foundation model from multimodal data and neuroscientific priors to capture universal propagation patterns. Stage 2 adapts this model to specific BCI scenarios such as emotion or motor imagery through fine-tuning. Stage 3 personalizes the output using subject-specific calibration for high-fidelity signals; with limited data zero-shot learning provides partial adaptation.}
    \label{fig:fig6}
\end{figure*}

Furthermore, this study proposes a three-stage evolutionary paradigm as a foundational roadmap for the future development of EEG signal enhancement, illustrated in Figure~\ref{fig:fig6}. The lower tier, Stage 1, establishes a neural representation foundation model based on “data plus knowledge,” which learns universal principles of neural signal propagation by integrating large-scale datasets of EEG, iEEG, and MEG with neuroscientific priors to compute shared representations of neural activity patterns, thereby enabling robust cross-scenario transfer and multi-dataset generalization. The middle tier, Stage 2, implements scenario-oriented fine-tuned adaptation: the neural representation foundation model is further fine-tuned on task-specific BCI datasets—such as those for affective computing or motor imagery—to yield a scenario-specific adaptation module. Integration of this module endows the enhancement framework with strong contextual adaptability, allowing it to generate EEG enhanced signals that are better aligned with the demands of a given application domain. The upper tier, Stage 3, focuses on personalized representation enhancement. Here, subject-specific calibration data are used to perform personalized fine-tuning of the Stage 2 module, enabling individualized modeling across participants and producing EEG enhanced signals with maximized fidelity—critical for advanced applications such as high-precision BCI control and targeted cortical activity reconstruction. When individualized data are scarce, zero-shot learning strategies can be employed to partially compensate and improve model performance.

This three-stage paradigm not only offers a systematic solution to current bottlenecks—including data scarcity, low signal-to-noise ratio, and task heterogeneity, but also marks a paradigm shift in non-invasive brain–computer interfaces from passive signal enhancement toward active neural information reconstruction. Pursuing this trajectory, future BCI systems are expected to achieve high fidelity, task adaptability, and individual interpretability while maintaining hardware lightweightness, thereby enabling robust deployment in critical real-world applications such as neurorehabilitation, human–machine interaction, and defense.

\section*{Methods}\label{sec4}
To overcome the fundamental limitations of EEG—particularly its substantially lower signal fidelity compared to iEEG, while avoiding the clinical inaccessibility of iEEG due to its invasive nature, this study proposes a unified EEG--iEEG representation enhancement framework that synergistically integrates data-driven learning with neuroscientific priors. The framework enhances raw EEG signals by jointly leveraging a pretrained large-scale EEG model, neural dynamics constraints, and brain geometric modeling. Its primary objective is to reconstruct the high-dimensional neural activity representations that are attenuated during propagation through the brain due to volume conduction and tissue filtering, thereby generating enhanced EEG signals whose fidelity approaches that of iEEG. As illustrated in Figure~1b, the pipeline comprises four coordinated stages: (i) extracting robust generic neural representations from raw EEG using a pretrained foundation model; (ii) computing constraint features over cortical regions based on neural dynamics equations; (iii) modeling the intrinsic geometric properties of the cortical surface via geometric mode decomposition; and (iv) progressively reconstructing high-fidelity enhanced EEG signals under the guidance of these three complementary feature streams, using a multi-dimensional EEG feature encoder coupled with a diffusion-based generative mechanism.

Critically, the output of this framework is neurophysiologically unified: when evaluated for iEEG reconstruction and signal fidelity, it is termed “synthesized iEEG”; when deployed in downstream BCI tasks, it is referred to as “enhanced EEG.” These two designations share an identical generative mechanism and representational basis, with the naming distinction reflecting only the objective of the downstream application, not any divergence in the signal’s intrinsic properties.

\subsection*{Pretrained EEG Representation Encoding Module}
Raw EEG signals are highly susceptible to contamination from non-neural artifacts such as electromyographic and electrooculographic activity. This results in low signal-to-noise ratio and high inter-subject variability, which impedes the reliable extraction of high-dimensional neural activity patterns. Even after standard preprocessing, signal fidelity remains substantially limited, making it difficult to support demanding downstream tasks such as high-precision neural decoding. A more fundamental challenge arises from the scarcity of simultaneously recorded EEG--iEEG paired data in clinical and research settings. Publicly available datasets typically involve only a small number of subjects, a narrow range of experimental paradigms, and short recording durations. Under such limited and low-quality observational conditions, reconstruction of high-dimensional, spatiotemporally continuous neural representations is prone to overfitting subject-specific noise or task-related artifacts, significantly compromising generalization to unseen scenarios.

To address these challenges, a Pretrained EEG Representation Encoder (PERE) is constructed. PERE employs LaBraM, a large foundation model pretrained on massive-scale unpaired EEG data, as its initial network layer to extract noise-robust and cross-task generic pretrained neural representations. These representations do not require iEEG supervision yet effectively disentangle neural activity from multi-source confounding factors, thereby establishing a foundational neural representation space for subsequent multi-dimensional feature fusion. Specifically, PERE consists of three components: an EEG spatial dimension adapter, a pretrained generic neural representation extractor, and a neural representation spatial aligner.

\subsubsection*{EEG Spatial Dimension Adapter}

Let a preprocessed EEG segment be denoted as $X_i \in \mathbb{R}^{C \times T}$, where $C$ is the number of channels and $T$ is the number of time points (fixed at 250 in this study). To map these signals into the embedding space of the pre-trained model, we introduce a spatial dimension adapter, a parameter-shared single-layer linear projection applied at each channel–time position:
\[
[X_i^{\mathrm{emb}}]_{c,t,:} = \mathbf{W}_{\mathrm{adapter}} \cdot X_i(c,t) + \mathbf{b}_{\mathrm{adapter}}.
\]

Here, $X_i(c,t) \in \mathbb{R}$ denotes a scalar input, and $[X_i^{\mathrm{emb}}]_{c,t,:} \in \mathbb{R}^{D_{\mathrm{emb}}}$ is the corresponding embedding vector. This operation is equivalent to generating $C \times T$ tokens on a $C \times T$ grid, with each token embedded in a $D_{\mathrm{emb}}$-dimensional feature space. The resulting tensor is given by
\[
X_i^{\mathrm{emb}} = \mathrm{Adapter}(X_i) \in \mathbb{R}^{C \times T \times D_{\mathrm{emb}}}.
\]

\subsubsection*{Pretrained Generic Neural Representation Extractor}

The tensor $X_i^{\mathrm{emb}} \in \mathbb{R}^{C \times T \times D_{\mathrm{emb}}}$ serves as input to LaBraM. LaBraM is a Transformer architecture pretrained on large-scale unpaired EEG data, containing 5.8 million parameters, with its self-attention mechanism computed over all $C \times T$ tokens:
\[
\mathrm{Attention}(\mathbf{Q}, \mathbf{K}, \mathbf{V}) = \mathrm{softmax}\!\left( \frac{\mathbf{Q} \mathbf{K}^\top}{\sqrt{d}} \right) \mathbf{V},
\]
where $d = D_{\mathrm{emb}}$ denotes the feature dimension. The core advantage of this mechanism lies in its ability to adaptively capture nonlinear spatiotemporal interactions prevalent in EEG, such as cross-frequency phase--amplitude coupling and event-related desynchronization/synchronization. These patterns constitute essential carriers of neural information encoding and cannot be effectively modeled by conventional bandpass filtering or handcrafted features.

The activations from the $L^\ast = 12$-th layer of LaBraM are selected as intermediate representations:
\[
E_i^{\mathrm{raw}} = \mathrm{LaBraM}_{L^\ast}(X_i^{\mathrm{emb}}) \in \mathbb{R}^{C \times T \times D_{\mathrm{emb}}}.
\]
Empirical evaluation shows that this layer achieves an optimal trade-off between preserving fine-grained spectral details and encoding high-level neural semantics. It consistently demonstrates the strongest generalization across datasets and subjects.

\subsubsection*{Neural Representation Spatial Alignment Module}

To align the output of \textsc{LaBraM} with the multi-dimensional EEG feature space required by subsequent fusion modules, a linear projection layer is applied to each token:
\[
[E_i^{\mathrm{proj}}]_{c,t,:} = \mathbf{W}_{\mathrm{proj}} \cdot [E_i^{\mathrm{raw}}]_{c,t,:} + \mathbf{b}_{\mathrm{proj}}.
\]

Here, $E_i^{\mathrm{raw}} \in \mathbb{R}^{C \times T \times D_{\mathrm{emb}}}$ denotes the raw output of LaBraM, and $D_{\mathrm{emb}}$ is its embedding dimension. The projected tensor $E_i^{\mathrm{proj}} \in \mathbb{R}^{C \times T \times D_{\mathrm{fuse}}}$ is then flattened and linearly compressed into a unified sequence space:
\[
\mathbf{E}_{\mathrm{data}} = \mathrm{Reshape}(E_i^{\mathrm{proj}}) \in \mathbb{R}^{L \times D},
\]
where $L = C \cdot T$ and $D = D_{\mathrm{fuse}}$. This yields the final data-driven pretrained EEG representation $\mathbf{E}_{\mathrm{data}}$. The resulting representation not only suppresses non-neural artifacts in EEG signals but also encodes their underlying neural activity patterns. It thus provides a robust, generalizable foundation for integrating neural dynamics and cortical geometric constraints, enabling the reconstruction of high-dimensional dynamic neural information lost during transcranial signal propagation.

\subsection*{Neural Dynamics Constraint Computation Module}

Although the data-driven representation $\mathbf{E}_{\mathrm{data}}$ exhibits strong generalization capability, its generation process does not explicitly adhere to the fundamental physical laws of neuroelectrophysiology. This may cause the enhanced EEG signals to deviate from genuine cortical activity in temporal dynamics or spatial distribution, thereby compromising their neuroscientific plausibility and the reliability of downstream decoding tasks. To ensure interpretability and fidelity at the level of neural mechanisms, a prior model grounded in neural dynamics is introduced to impose physical constraints consistent with established neural signal propagation principles during signal generation. This module comprises three coordinated submodules: (i) head physical modeling, (ii) neural population dynamics representation, and (iii) dynamics-to-embedding mapping.

\subsubsection*{Head Physical Modeling Submodule}

Specifically, a three-layer Boundary Element Method (BEM) head model is constructed from individual T1-weighted MRI scans, corresponding respectively to brain tissue, skull, and scalp. Within this model, the relationship between scalp potentials $\Phi_k$ and cortical current sources $\mathbf{J}$ is governed by the Poisson equation derived from the quasi-static Maxwell equations:
\[
\nabla \cdot (\sigma_k \nabla \Phi_k) = 0 \quad \text{in } \Omega_k,
\]
where $\Omega_k$ denotes the $k$-th tissue compartment ($k = 1$: brain, $k = 2$: skull, $k = 3$: scalp), and the conductivity vector is $\boldsymbol{\sigma} = [0.33,\, 0.01,\, 0.33]^\top~\mathrm{S/m}$. This formulation accurately captures the strong attenuation and spatial low-pass filtering effect of the skull on high-frequency neural activity. Such physics-mediated distortion—arising from the multi-layered anatomical structure of the head—is the primary cause of the loss of fine-grained neural patterns in raw EEG.

\subsubsection*{Neural Population Dynamics Representation Submodule}

Under this geometric constraint, the lead field matrix $\mathbf{L} \in \mathbb{R}^{C \times N}$ is computed as:
\[
\mathbf{L} = \mathbf{G} \mathbf{B},
\]
where $\mathbf{G}$ denotes the Green’s function and $\mathbf{B}$ is the mapping operator; $N$ is the number of cortical source points. Given the $i$-th scalp EEG segment $\mathbf{X}_i \in \mathbb{R}^{C \times T}$, the cortical current source estimate $\hat{\mathbf{J}}_i \in \mathbb{R}^{N \times T}$ is obtained by solving the sLORETA-regularized optimization problem:
\[
\hat{\mathbf{J}}_i = \arg\min_{\mathbf{J}} \left\| \mathbf{W}^{-1/2} (\mathbf{X}_i - \mathbf{L} \mathbf{J}) \right\|_F^2 + \lambda \left\| \mathbf{J} \right\|_F^2,
\]
where $\mathbf{W} \in \mathbb{R}^{C \times C}$ is a noise-covariance-based normalization matrix, $\lambda > 0$ is the regularization parameter, and $\|\cdot\|_F$ denotes the Frobenius norm. This strategy ensures solution uniqueness while preserving the true spatial distribution of cortical activity, thereby providing a reliable constraint $\hat{\mathbf{J}}_i$.

To further constrain temporal dynamics, $\hat{\mathbf{J}}_i(\mathbf{x}, t)$ serves as the external input to a Neural Mass Model (NMM) that describes the mean membrane potential evolution of local excitatory ($E$) and inhibitory ($I$) neuronal populations:
\[
\begin{cases}
\dot{V}_E = f_E(V_E, V_I, \hat{J}), \\
\dot{V}_I = f_I(V_E, V_I, \hat{J}),
\end{cases}
\]
where $f_E$ and $f_I$ are nonlinear dynamical functions simplified from Hodgkin--Huxley theory. This model effectively captures characteristic oscillatory rhythms—such as alpha and beta bands—generated by cortical microcircuits.

Let $\Delta V(\mathbf{x}, t)$ denote the deviation of the excitatory population’s mean membrane potential from rest at cortical location $\mathbf{x} \in \Omega$, where $\Omega$ is the subject-specific cortical surface manifold. Similarly, $\mathbf{y} \in \Omega$ denotes an integration variable representing other cortical locations. The spatiotemporal evolution of $\Delta V(\mathbf{x}, t)$ is then governed by the neural field equation:
\[
\frac{\partial \Delta V(\mathbf{x}, t)}{\partial t} =-\Delta V(\mathbf{x}, t) +
\]
\[
\int_{\Omega} w\big(d_{\mathrm{geo}}(\mathbf{x}, \mathbf{y})\big) \, S\big(\Delta V(\mathbf{y}, t)\big) \, \mathrm{d}\mathbf{y} + \hat{J}(\mathbf{x}, t)
\]

$w(\cdot)$ is a connectivity kernel based on geodesic cortical distance $d_{\mathrm{geo}}(\mathbf{x}, \mathbf{y})$, quantifying structural coupling strength between regions;
$S(\cdot)$ is a sigmoidal activation function;
$\hat{J}(\mathbf{x}, t)$ is the cortical current source estimated via sLORETA.

This equation explicitly models the diffusion and synchronization of neural activity across the cortical surface, revealing how neural signals evolve in space and time during transcranial propagation.

\subsubsection*{Neural Population Dynamics Representation Mapping Submodule}

Finally, to interface the computed spatiotemporal neural activity dynamics with downstream deep generative modules, a lightweight multilayer perceptron (MLP) compresses the high-dimensional dynamical state $\Delta V(\mathbf{x}, t)$ into a compact prior feature:
\[
\mathbf{E}_{\mathrm{prior}} = \mathrm{MLP}\big(\mathrm{flatten}(\Delta V)\big) \in \mathbb{R}^{L \times D},
\]
where $\mathrm{flatten}(\Delta V)$ reshapes the cortical field $\Delta V(\mathbf{x}, t)$ over space $\mathbf{x} \in \Omega$ and time $t$ into a vector of length $L$, and $D$ denotes the embedding dimension. This mapping preserves essential dynamic neural representations while enabling a compatible transformation from the continuous differential system to a discrete deep learning representation. Consequently, the derived neural dynamics constraint can be effectively fused into the subsequent diffusion-based enhancement process.

\subsection*{Geometric coding of the cerebral cortex}

Although the Boundary Element Method (BEM) model effectively captures the volume conduction effect from cortex to scalp, distortions in raw EEG also arise from the spatial organization of cortical neural activity on the subject-specific cortical manifold. Inter-subject variability in sulcal and gyral morphology modulates local neural synchrony and the spatial superposition of field potentials, leading to significant inter-individual differences in scalp EEG even under identical cognitive states. Such anatomy-coupled neural representations are further obscured during transcranial propagation and cannot be directly recovered from observed signals. To address this, a structural modeling pipeline is constructed, comprising three coordinated submodules: (i) cortical surface geometric decomposition, (ii) geometric mode feature computation, and (iii) weighted geometric feature mapping.

\subsubsection*{Cortical Surface Geometric Decomposition Submodule}

To recover neural activity patterns tightly coupled with cortical geometry, a geometric mode decomposition framework is adopted. Specifically, a high-resolution cortical triangular mesh $\mathcal{M}$ is reconstructed from individual T1-weighted MRI using FreeSurfer, and the Laplace--Beltrami Operator (LBO) is defined on $\mathcal{M}$ as:
\[
\Delta_{\mathcal{M}} f = \frac{1}{\sqrt{\det g}} \partial_i \left( \sqrt{\det g} \, g^{ij} \partial_j f \right),
\]
where $g$ denotes the Riemannian metric tensor induced by the mesh. Solving the eigenvalue problem:
\[
\Delta_{\mathcal{M}} \phi_k = -\lambda_k \phi_k, \qquad k = 0, 1, 2, \dots,
\]
yields an orthogonal set of geometric modes $\{\phi_k\}$ on $\mathcal{M}$, with corresponding eigenvalues $\lambda_k$ reflecting spatial frequency (i.e., oscillation scale). These modes form an intrinsic eigenfunction basis of the cortical surface, naturally encoding geometric properties such as gyral scale, curvature variation, and topological connectivity.

\subsubsection*{Geometric Mode Feature Computation Submodule}

To capture the multi-scale modulation of neural activity by cortical geometry, the spectral domain is partitioned into three bands: low ($\lambda_k < \lambda_{\mathrm{low}}$), medium ($\lambda_{\mathrm{low}} \leq \lambda_k < \lambda_{\mathrm{high}}$), and high ($\lambda_k \geq \lambda_{\mathrm{high}}$). The T1-weighted image intensity $I_{\mathrm{T1}}(\mathbf{r})$ is then projected onto the geometric eigenbasis:
\[
\alpha_k = \langle I_{\mathrm{T1}}, \phi_k \rangle_{\mathcal{M}} = \int_{\mathcal{M}} I_{\mathrm{T1}}(\mathbf{r}) \, \phi_k(\mathbf{r}) \, \mathrm{d}\mathbf{r}.
\]
This projection is neurobiologically grounded: T1 signal intensity correlates strongly with cortical myelination, neuronal density, and microcolumnar architecture—microstructural features that govern the generation efficiency and spatial coherence of local field potentials. By representing T1 MRI in the geometric eigenbasis, microscopic cytoarchitectonic information is unified with macroscopic cortical morphology within a common spectral framework, yielding a geometric--functional joint representation that is both anatomically precise and physiologically meaningful.

\subsubsection*{Weighted Geometric Feature Mapping Submodule}

Finally, a geometric prior feature map is reconstructed via learnable band-wise weights $\mathbf{w} = [w_{\mathrm{low}}, w_{\mathrm{mid}}, w_{\mathrm{high}}]^\top$:
\[
\mathbf{E}_{\mathrm{geo}}(\mathbf{r}) = \sum_{k} w_b(\lambda_k) \, \alpha_k \, \phi_k(\mathbf{r}),
\]
where $w_b(\lambda_k)$ equals $w_{\mathrm{low}}$, $w_{\mathrm{mid}}$, or $w_{\mathrm{high}}$ depending on the band to which $\lambda_k$ belongs. The weight $w_b$ reflects the detectability of each spatial frequency component in scalp EEG. This feature map explicitly encodes how individual cortical geometry influences neural signal propagation, thereby providing critical geometric guidance for the enhancement process. As a result, the generated enhanced EEG signals conform not only to neurophysiological dynamics but also to the subject’s true cortical geometry in spatial distribution.

To facilitate fusion with the data-driven representation, $\mathbf{E}_{\mathrm{geo}}$ is resampled to match the spatiotemporal grid of $\mathbf{E}_{\mathrm{data}}$ and mapped into the shared feature space $\mathbb{R}^{L \times D}$ via a lightweight projection layer, yielding $\mathbf{E}_{\mathrm{struct}}$. This structural feature serves as a conditional input during EEG enhancement, ensuring that neural activity patterns masked by the transcranial propagation pathway are reconstructed under explicit, individualized brain geometric constraints.

\subsection*{EEG Enhancement Signal Generation via Feature Fusion and Diffusion Modeling}

\subsubsection*{Multi-Dimensional Data--Knowledge Feature Fusion Module}

To synergistically integrate the three complementary representations—data-driven generic neural features $\mathbf{E}_{\mathrm{data}}$, neural dynamics constraint prior $\mathbf{E}_{\mathrm{prior}}$, and cortical geometric structure prior $\mathbf{E}_{\mathrm{struct}}$—a multi-dimensional data--knowledge feature fusion module is designed. Its objective is to construct a conditional signal that guides the enhancement process to simultaneously satisfy statistical adaptability, neurophysiological plausibility, and anatomical specificity.

The fusion module concatenates the three feature streams $\mathbf{E}_{\mathrm{data}}$, $\mathbf{E}_{\mathrm{prior}}$, and $\mathbf{E}_{\mathrm{struct}}$ along the sequence dimension to form a unified input, which is then processed by a shallow Transformer encoder. Cross-modal dependencies are modeled via multi-head self-attention:
\[
\mathbf{Z}_{\mathrm{fused}} = \mathrm{TransEncoder}\big([\mathbf{E}_{\mathrm{data}}; \mathbf{E}_{\mathrm{prior}}; \mathbf{E}_{\mathrm{struct}}]\big).
\]
This architecture dynamically weights the contribution of each feature stream in representing neural activity patterns, thereby avoiding semantic conflicts arising from rigid concatenation. The resulting fused representation $\mathbf{Z}_{\mathrm{fused}}$ preserves the flexibility of data-driven learning while embedding the rigidity of neurodynamical and geometric constraints.

\subsubsection*{High-Fidelity EEG Enhancement via Fusion-Guided Diffusion}

The fused representation $\mathbf{Z}_{\mathrm{f}}$ serves as the global conditioning signal for a conditional diffusion model, which generates an enhanced EEG signal $\widetilde{X} \in \mathbb{R}^{C \times T}$ with fidelity approaching that of iEEG. The enhancement process is formulated as a progressive denoising trajectory from pure Gaussian noise $X_T \sim \mathcal{N}(0, I)$ to the target signal $X_0 = \widetilde{X}$. The reverse Markov chain is defined as:
\[
p_\theta(X_{t-1} \mid X_t, \mathbf{Z}_{\mathrm{f}}) = \mathcal{N}\big(X_{t-1}; \mu_\theta(X_t, t, \mathbf{Z}_{\mathrm{f}}), \Sigma_\theta(t)\big),
\]
where $\mu_\theta$ is implemented by a U-Net architecture. Skip connections in the U-Net preserve fine-grained temporal details from the original EEG, while the global condition $\mathbf{Z}_{\mathrm{fused}}$ is injected at every decoder layer via cross-attention mechanisms, ensuring consistency between the generated signal and the multi-dimensional guidance features across multiple scales.

Notably, during early diffusion steps (large $t$), the model prioritizes recovery of high-frequency neural patterns masked by the skull’s low-pass filtering effect. In later steps (small $t$), it refines phase and amplitude to achieve precise temporal alignment with the observed EEG. This staged, multi-scale control enables the model to learn both global neural event structures and high-dimensional spatiotemporal dynamics aligned with iEEG, thereby producing high-fidelity enhanced EEG signals.

\subsubsection*{Composite Loss Function}

The model is trained end-to-end by minimizing the variational Evidence Lower Bound (ELBO):

\[
\mathcal{L}_{\mathrm{diff}} = \mathbb{E}_{t, X_0, \epsilon}
\]
\[
\left[ \left\| \epsilon - \epsilon_\theta\big( \sqrt{\alpha_t} X_0 + \sqrt{1 - \alpha_t^2} \, \epsilon,\, t,\, \mathbf{Z}_{\mathrm{fused}} \big) \right\|_2^2 \right], 
\]

where $\epsilon \sim \mathcal{N}(0, I)$ denotes standard Gaussian noise and $X_t = \sqrt{\alpha_t} X_0 + \sqrt{1 - \alpha_t^2} \, \epsilon$ is the forward noising process. This objective enforces distributional consistency between the generated signal $\widetilde{X}$ and the three guiding feature streams, thereby ensuring generalization across diverse non-invasive BCI scenarios.

To further improve amplitude and waveform fidelity, three auxiliary supervised losses are incorporated into the training objective:
\begin{itemize}
    \item A weighted negative Pearson correlation loss $\lambda_1 \cdot \big(-\rho(\hat{X}_0, X_0)\big)$ to enhance morphological similarity between the generated and target waveforms;
    \item A mean squared error term $\lambda_2 \cdot \|\hat{X}_0 - X_0\|_2^2$ to enforce precise amplitude reconstruction;
    \item An $L_2$ weight regularization term $\lambda_3 \cdot \|\theta\|_2^2$ to mitigate overfitting.
\end{itemize}
The total loss is defined as:
\[
\mathcal{L}_{\mathrm{total}} = 
\]
\[
\mathcal{L}_{\mathrm{diff}} + \lambda_1 \big(-\rho(\hat{X}_0, X_0)\big) + \lambda_2 \|\hat{X}_0 - X_0\|_2^2 + \lambda_3 \|\theta\|_2^2.
\]

In this study, the hyperparameters are set to $\lambda_1 = 0.1$, $\lambda_2 = 0.3$, and $\lambda_3 = 0.1$. This composite objective function jointly optimizes the generation process across four dimensions—distributional consistency, amplitude accuracy, waveform morphology, and model complexity—ensuring that the generated enhanced EEG signal $\widetilde{X}$ achieves comprehensive fidelity approaching that of iEEG.

\subsection*{Downstream Task Evaluation Pipeline for EEG Datasets}

To systematically evaluate the performance of enhanced EEG signals across diverse downstream tasks, a Transformer-based autoencoder module is constructed to adapt conventional EEG signals to the proposed enhancement framework, enabling spatial channel compression and subsequent reconstruction.

\subsubsection*{EEG Spatial Channel Compression Module}

The encoder takes raw EEG signals $\mathbf{X} \in \mathbb{R}^{64 \times T}$, recorded from standard 64-channel systems with temporal length $T$, as input. It models inter-channel spatial dependencies via multi-head self-attention and outputs a compressed representation $\mathbf{X}_{\mathrm{comp}} \in \mathbb{R}^{C \times T}$. The encoder is trained to reconstruct the subset of $C$ scalp EEG channels that spatially correspond to the intracranial electrode locations in simultaneously recorded EEG--iEEG datasets. This ensures compatibility between the compressed signal and the input requirements of the proposed enhancement framework.

\subsubsection*{EEG Enhancement via Unified Framework}

The compressed representation $\mathbf{X}_{\mathrm{comp}}$ is fed into the unified enhancement framework. Through sequential stages—including data-driven representation extraction, neurodynamical--geometric constraint fusion, and conditional diffusion-based generation—the framework produces a high-fidelity enhanced signal ${\widetilde{\mathbf{X}}}_{\mathrm{enh}} \in \mathbb{R}^{C' \times T}$. Here, the number of output channels $C'$ strictly matches the count of valid iEEG channels in the corresponding simultaneous recording, ensuring dimensional alignment with ground-truth iEEG in the channel space.

\subsubsection*{EEG Spatial Resolution Enhancement Module}
The enhanced signal ${\widetilde{\mathbf{X}}}_{\mathrm{enh}}$ is then passed to a decoder, which also adopts a Transformer architecture. During training, the decoder takes the $C'$ iEEG-proximal scalp channels as input and reconstructs the full 64-channel scalp EEG as the target. This enables the decoder to learn a mapping that effectively upsamples the enhanced signal back to full-brain spatial coverage, thereby completing end-to-end spatial resolution enhancement.

\subsubsection*{Downstream Task Experimental Protocol}

To ensure an objective assessment of the proposed framework, the performance of original EEG signals and generated enhanced EEG signals is compared under strictly controlled experimental conditions. Both signal types undergo identical preprocessing pipelines, temporal windowing strategies, dataset partitioning schemes, and hyperparameter search spaces, guaranteeing a fair evaluation. Downstream decoders employ simple yet well-established baseline architectures, including multilayer perceptrons (MLPs) and lightweight Transformers. These models possess sufficient representational capacity to capture fundamental neural dynamics while avoiding excessive complexity that could obscure the true impact of signal fidelity differences on task performance. Across twelve EEG datasets spanning multiple task types, systematic experiments are conducted using identical decoders for both signal variants. The consistent performance gains observed in the enhanced signals validate the effectiveness of the proposed framework in improving the fidelity of non-invasive EEG and boosting decoding accuracy in downstream BCI applications.

\subsection*{Data preprocessing}
All neurophysiological and neuroimaging data underwent standardized preprocessing pipelines to ensure methodological rigor and cross-dataset comparability.

For simultaneously recorded EEG and iEEG signals, raw data were bandpass filtered between 0.5 and 150 Hz to remove baseline drift and high-frequency artifacts, followed by a 50 Hz notch filter to suppress line noise. iEEG signals were re-referenced using a bipolar montage to reduce common-mode noise, and channels with signal-to-noise ratio below 3 dB or impedance exceeding 20 kΩ were manually excluded. EEG signals were further processed using independent component analysis (ICA) combined with the automatic artifact rejection algorithm ADJUST to remove ocular, cardiac, and muscular artifacts; only channels with signal-to-noise ratio above 3 dB and impedance below 10 kΩ were retained. Both signal types were re-referenced to the common average, downsampled uniformly to 500 Hz, and segmented into non-overlapping epochs of 250 samples (corresponding to 500 ms). Each channel–epoch unit was independently z-score normalized, yielding standardized input tensors of dimension \(\mathbb{R}^{C \times 250}\), where \(C\) denotes the number of valid channels.

The twelve multi-task EEG datasets were processed using the same filtering pipeline, with additional ocular artifact correction via linear regression and electromyographic interference suppression through wavelet-enhanced ICA. After common average referencing, epochs exhibiting variance exceeding five times the median absolute deviation across the cohort were discarded. All signals were downsampled to 500 Hz, segmented into non-overlapping 250-sample windows, and z-score normalized per channel–epoch unit to produce \(\mathbb{R}^{C \times 250}\) tensors.

Structural MRI data were processed using FreeSurfer v7.3.0. T1-weighted images underwent motion correction, skull stripping, and N3 bias field correction, followed by intensity normalization and white matter segmentation. Cortical surface reconstruction employed deformable models expanded to the gray–white and pial boundaries, then smoothed with a 10 mm FWHM Gaussian kernel. All surface meshes were registered to the fsaverage5 standard space, and vertex-wise cortical thickness and curvature features were extracted and bilinearly interpolated onto a 642-vertex spherical grid. Quality control followed ENIGMA consortium guidelines: samples with surface overlap error greater than 0.5 mm or thickness signal-to-noise ratio below 8 dB were excluded. The resulting cortical geometric features were strictly aligned with electrophysiological data in individual anatomical space.

All preprocessing steps were implemented using MNE-Python 1.4 and FreeSurfer 7.3.0, encapsulated in containerized workflows to ensure cross-platform reproducibility. Processed datasets were partitioned into non-overlapping training and testing sets at an 8:2 ratio to prevent data leakage.

\backmatter

\bibliographystyle{sn-nature.bst}

\bibliography{sn-bibliography}




\setcounter{table}{0}
\renewcommand{\tablename}{}
\renewcommand{\thetable}{Extended Data Table \arabic{table}}

\setcounter{figure}{0}
\renewcommand{\figurename}{Extended Data Fig.}  
\renewcommand{\thefigure}{\arabic{figure}}  

\section*{Supplement}\label{sec11}

\begin{figure*}[htbp]
    \centering
    \includegraphics[width=\linewidth]{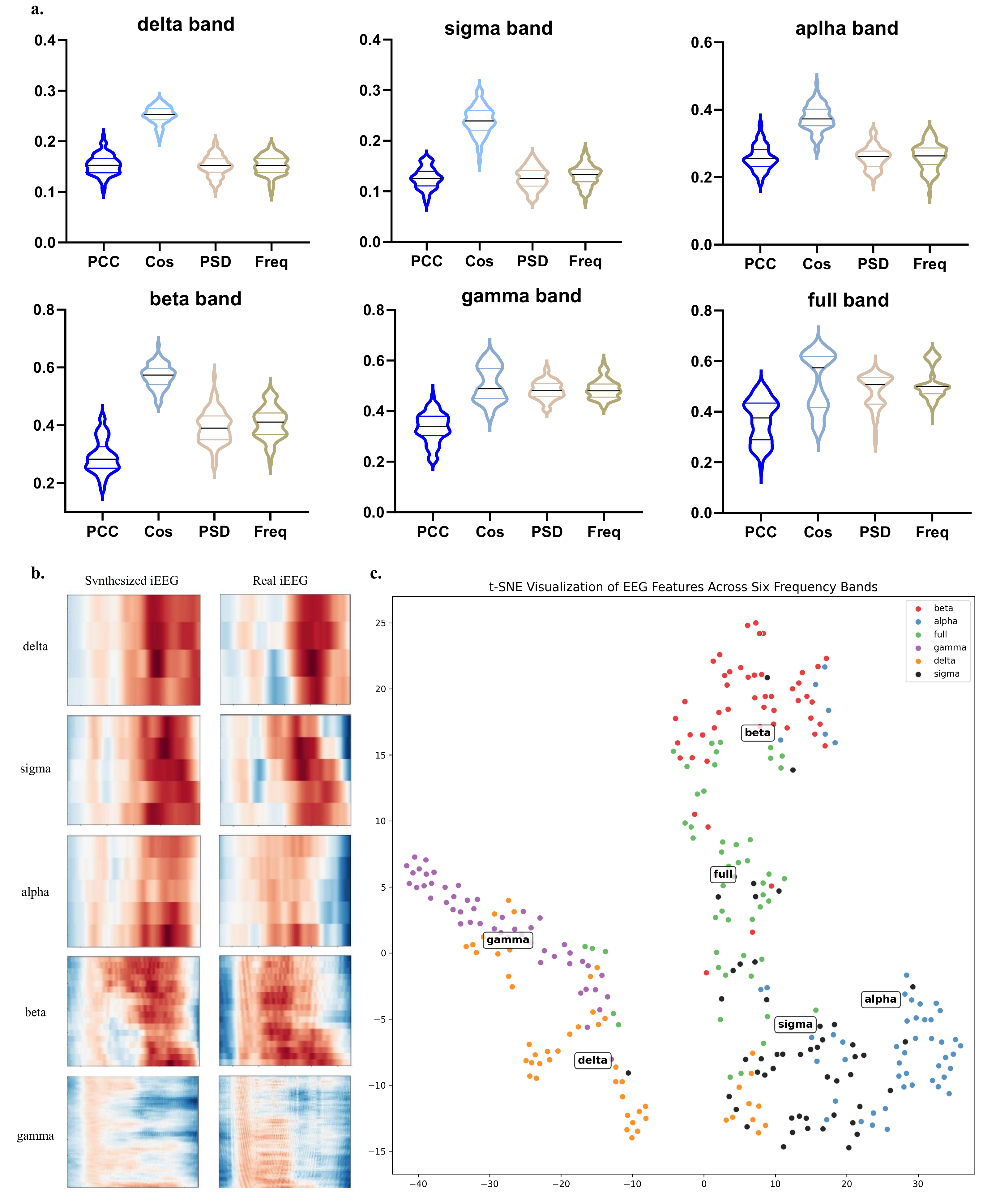}
    \caption{Spectral specificity of EEG enhancement across neural oscillation bands. a, Reconstruction fidelity increases with frequency; gamma achieves highest performance with PCC 0.47 and PSD Sim 0.49 while delta and theta are lowest with PCC 0.17 and PSD Sim 0.18. b, Beta and gamma enhanced signals replicate focal iEEG-like patterns whereas delta and theta remain diffuse lacking structural specificity. c, t-SNE shows distinct band-specific clusters with adjacent bands spatially proximate; full-band embeds near beta and gamma and sigma from 13 to 15 hertz lies at the alpha–beta boundary consistent with its transitional role.}
    \label{fig:fig7}
\end{figure*}

To quantify the contribution of distinct neural oscillatory bands to EEG signal enhancement, this experimental set systematically evaluates the spectral-specific performance of the five standard frequency bands—delta, theta, alpha, beta, and gamma—within the proposed framework, encompassing both reconstruction fidelity and latent representational structure. On two simultaneously recorded EEG--iEEG datasets (Dataset1 and Dataset2 of data availability), raw EEG signals were processed through standard bandpass filters to isolate individual sub-band components. Under identical experimental conditions, each sub-band signal was fed into the proposed framework to generate synthesized iEEG, which was then quantitatively compared against ground-truth iEEG using multiple evaluation metrics. The results are presented in extended data Fig.~\ref{fig:fig7}.

Reconstruction performance exhibits a monotonic increase with rising frequency. The delta and theta bands achieve the lowest scores across all metrics, with a mean PCC of 0.17 and PSD Sim of 0.18, reflecting the limited capacity of low-frequency components to recover high-fidelity neural dynamics. In contrast, the gamma band yields the best performance, achieving a mean PCC of 0.47 and PSD Sim of 0.49, approaching the performance of full-band input. These findings indicate that high-frequency components carry richer reconstructible information within the current framework. Furthermore, the multi-scale feature encoding mechanism effectively compensates for the attenuation of high-dimensional neural activity representations during propagation through the brain.

Figure~ further evaluates the enhancement efficacy at the level of spectral structure. Heatmap visualizations depict the information distribution of enhanced signals and real iEEG across frequency bands, where red indicates regions of high information concentration and blue indicates regions of low concentration. The results show that in the beta and gamma bands, enhanced signals exhibit localized, high-intensity information clustering patterns that closely resemble those of real iEEG, demonstrating strong spatial correspondence. In contrast, in the delta and theta bands, enhanced signals display diffuse energy distributions with poorly defined boundaries, lacking the structural characteristics observed in real iEEG. This observation aligns with the quantitative trends and confirms that the proposed framework effectively reconstructs the frequency-domain organizational signatures intrinsic to iEEG in high-frequency bands, thereby preserving the spatial specificity required for physiological plausibility.

Figure~7c visualizes the latent representations of the six frequency-band signals using t-SNE dimensionality reduction, revealing their intrinsic structures in the embedding space. Feature points from distinct frequency bands form relatively independent clusters, indicating that the framework successfully preserves the spectral identity of each band during encoding. Adjacent bands—such as delta and theta, alpha and beta, and gamma and full-band—exhibit spatial proximity and partial overlap, reflecting the neurodynamical continuity and functional associations across the frequency spectrum. The representation of the full-band signal resides between multiple single-band clusters and shows substantial overlap with the gamma and beta clusters, consistent with its nature as a broadband composite signal. Additionally, the sigma component (13–15 Hz) localizes at the boundary between the alpha and beta clusters, aligning with its established physiological role as a transitional rhythm in sensorimotor processing.

\begin{figure*}[htbp]
    \centering
    \includegraphics[width=\linewidth]{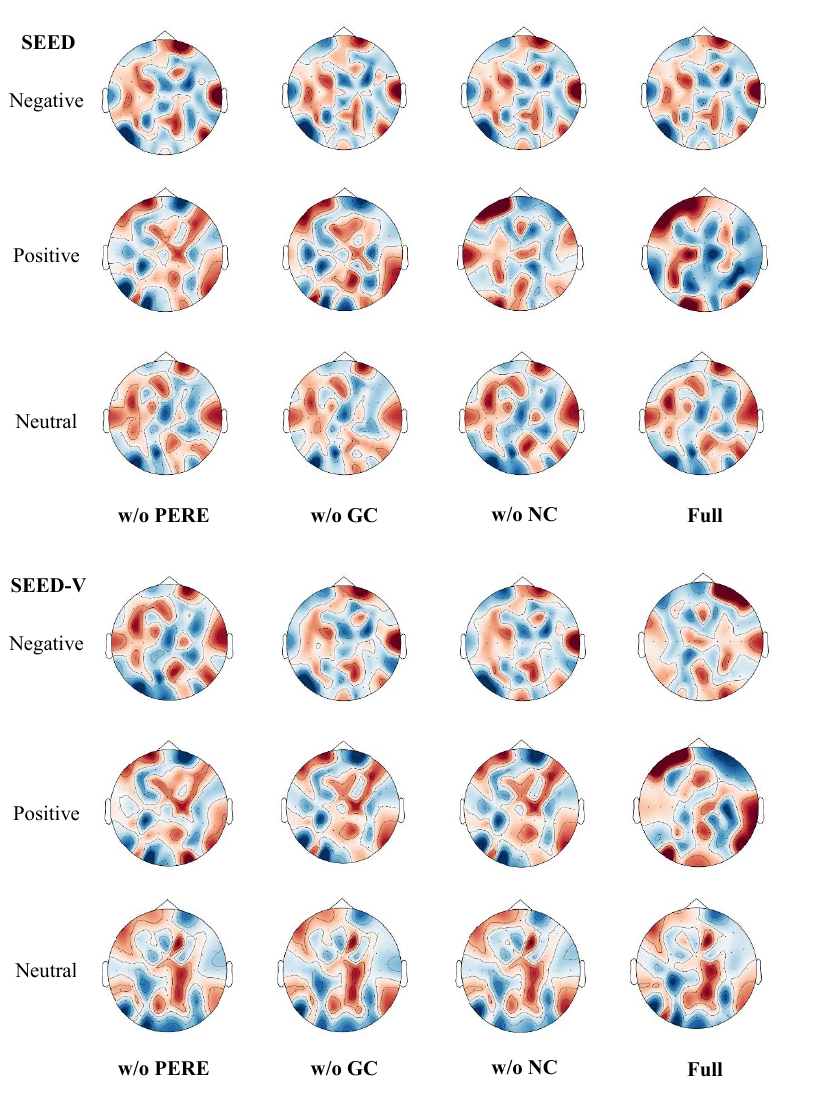}
    \caption{Topographic maps of EEG signal enhancement across model variants, illustrating the spatial specificity of neural activation patterns.Top rows show results on the SEED dataset; bottom rows show results on SEED-V. Columns correspond to three ablation settings: without the Pretrained EEG Representation Encoder (PERE), without the Neural Dynamics Constraint (NC), without the Geometric Constraint (GC), and the full model. Warm colors from red to orange indicate regions of high activation, while cool colors from blue to cyan denote low or baseline-level activity. As model components are incrementally integrated from ablated variants to the full framework, activation patterns become increasingly focal and anatomically coherent, converging on emotion-relevant cortical regions such as the prefrontal cortex, anterior cingulate, and temporal poles, while non-specific activations in sensorimotor and occipital areas are markedly suppressed. These findings demonstrate that the proposed framework enhances both the neurophysiological fidelity and functional specificity of reconstructed EEG signals.}
    \label{fig:sub2}
\end{figure*}

\begin{figure*}[htbp]
    \centering
    \includegraphics[scale=0.93]{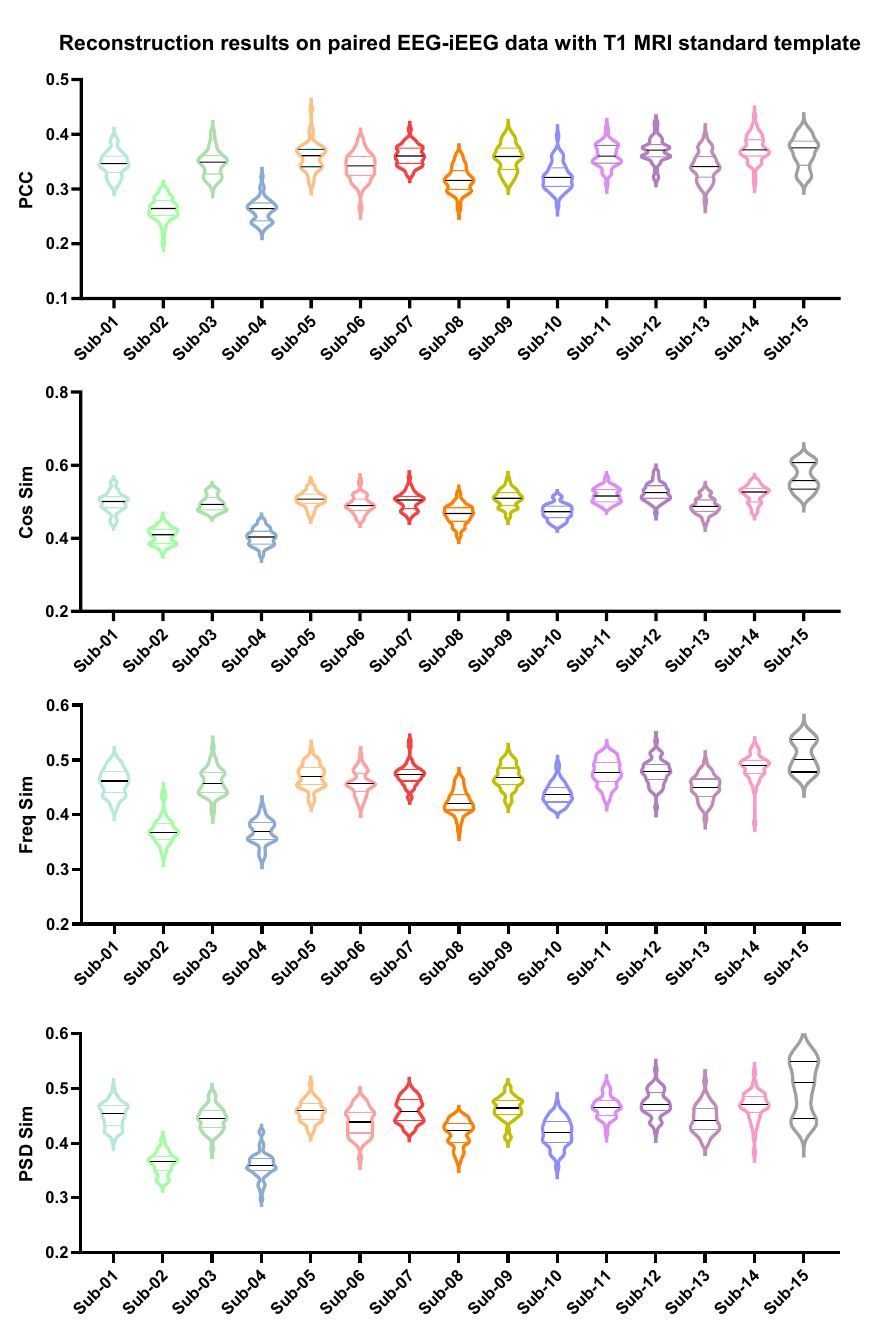}
    \caption{Reconstruction performance of iEEG signals on a T1 MRI–free EEG–iEEG dataset (OpenNeuro ds004752), demonstrating the generalizability and robustness of the proposed framework.
In the absence of subject-specific T1 MRI, the framework employs a standardized T1 MRI template as the anatomical prior to condition the iEEG reconstruction. Results are presented for 15 participants from dataset ds004752 (OpenNeuro, https://openneuro.org/datasets/ds004752/versions/1.0.1). Quantitative evaluation is conducted using four complementary metrics: Pearson correlation coefficient (PCC), cosine similarity (Cos Sim), power spectral density similarity (PSD Sim), and frequency-domain similarity (Freq Sim). The consistent high performance across these metrics confirms that the framework maintains efficacy and functional fidelity even when subject-specific structural imaging is unavailable, thereby validating its applicability in real-world clinical and research settings where high-resolution MRI may be inaccessible.}
    \label{fig:sub3}
\end{figure*}

\begin{figure*}[htbp]
    \centering
    \includegraphics[scale=0.3]{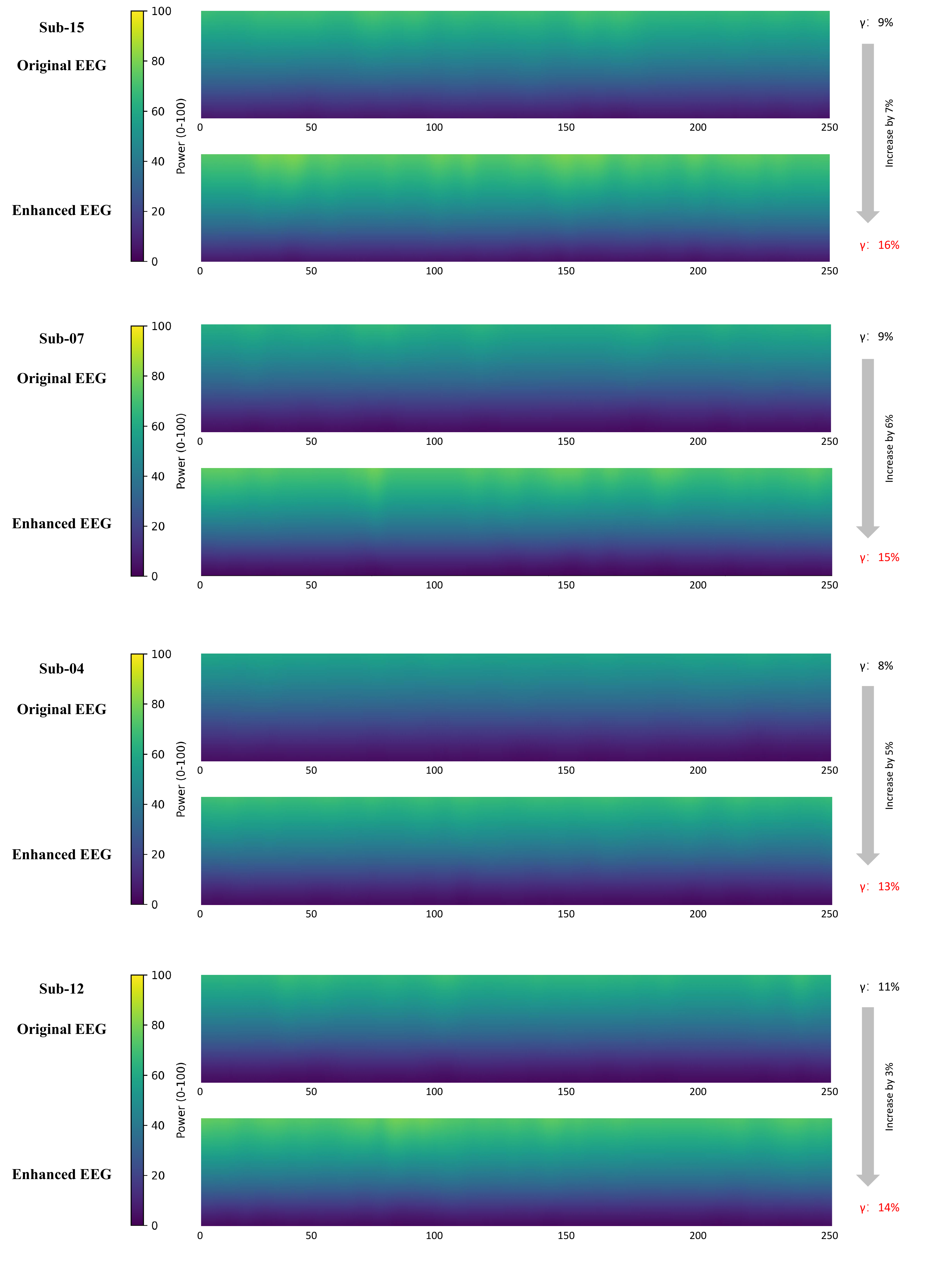}
    \caption{Comparative analysis of frequency distributions for original and enhanced EEG signals across representative subjects. The spectrograms demonstrate that the enhanced EEG signals exhibit a marked increase in the proportion of high-frequency bands (yellow-green hues) and a corresponding decrease in low-frequency bands (blue-purple hues) compared to the original signals. This spectral shift indicates a significant improvement in the proportion of high-frequency components, suggesting that the enhancement framework successfully and effectively recovers high-frequency information typically lost during the propagation of neural signals through brain tissue. Quantitative evaluation of the enhancement effects across different subjects reveals a widespread distribution of efficacy, with the proportion of increase in the $\gamma$ band following the order: Sub-15 (7\%) $>$ Sub-07 (6\%) $>$ Sub-04 (5\%) $>$ Sub-12 (3\%). This hierarchical trend highlights the robustness of the method in recovering high-frequency neural dynamics across varying individual baselines.}
    \label{fig:sup4}
\end{figure*}

\begin{figure*}[htbp]
    \centering
    \includegraphics[scale=0.27]{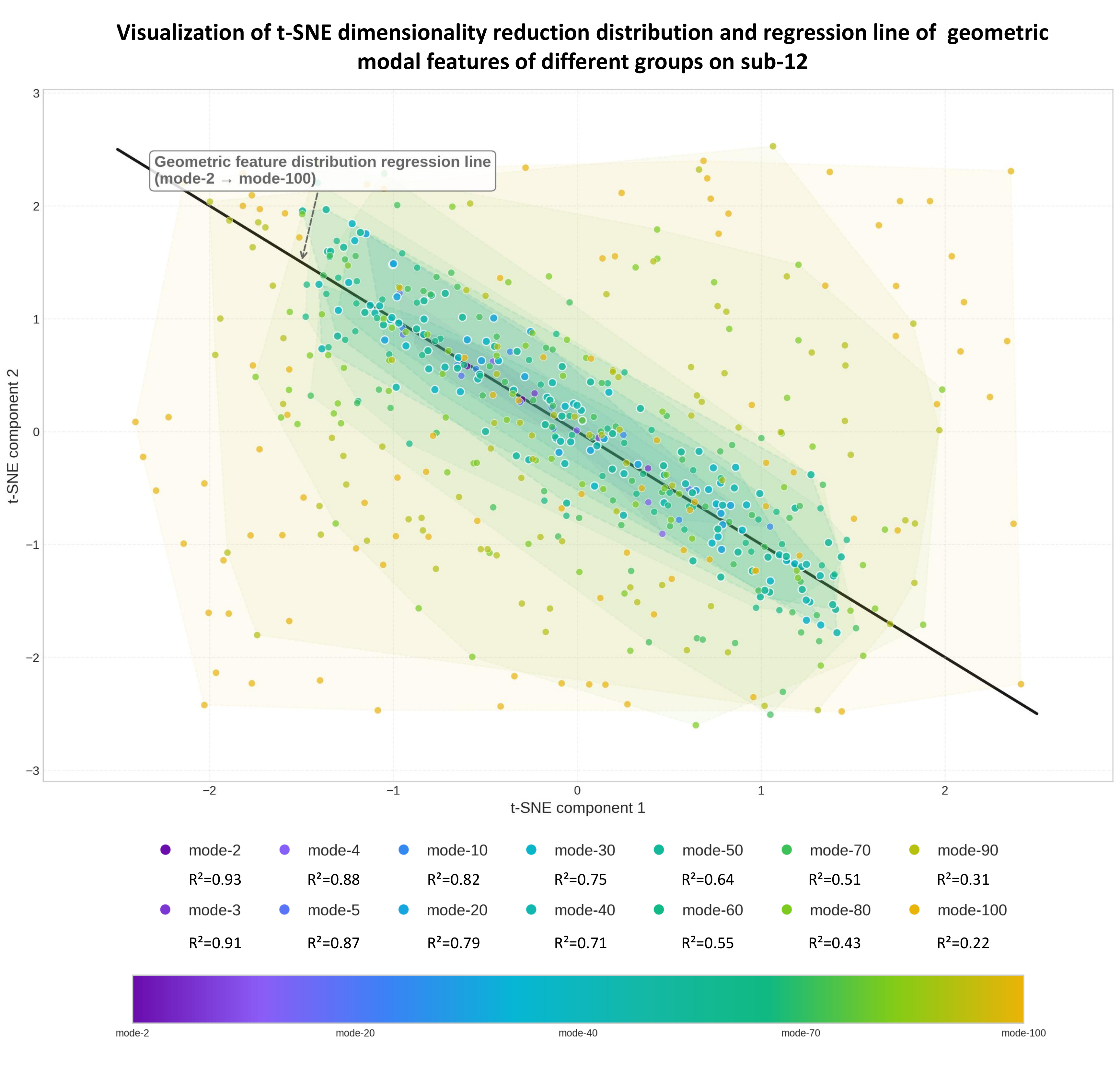}
    \caption{
       Scatter points represent geometric modal feature samples for mode groups (mode-2 to mode-100), color-coded with a purple$\rightarrow$blue$\rightarrow$green$\rightarrow$yellow gradient reflecting the progressive diffusion stage of geometric feature decomposition. The dotted black line represents the principal regression trajectory of the Laplace--Beltrami Operator, serving as a reference axis for quantifying geometric structure deviation, with R² values measuring the goodness-of-fit between each mode group and this regression line. Results show: mode-2 to mode-10 (purple-blue) exhibit highly compact clustering along the diagonal band with strong geometric constraint but limited discriminability; mode-60 to mode-100 (green-yellow) show excessive diffusion beyond the band structure, obscuring geometric-topological relationships; mode-20 to mode-50 (blue-green) maintain band-aligned structure with sufficient inter-sample discriminability while remaining tightly constrained around the regression line, preserving intrinsic cortical manifold geometry. These findings indicate that mode-20 to mode-50 achieve optimal balance between geometric constraint and feature expressiveness, representing the most suitable geometric feature decomposition modes for EEG enhancement tasks.
    }
    \label{fig:sup51}
\end{figure*}

\begin{figure*}[htbp]
    \centering
    \includegraphics[scale=0.27]{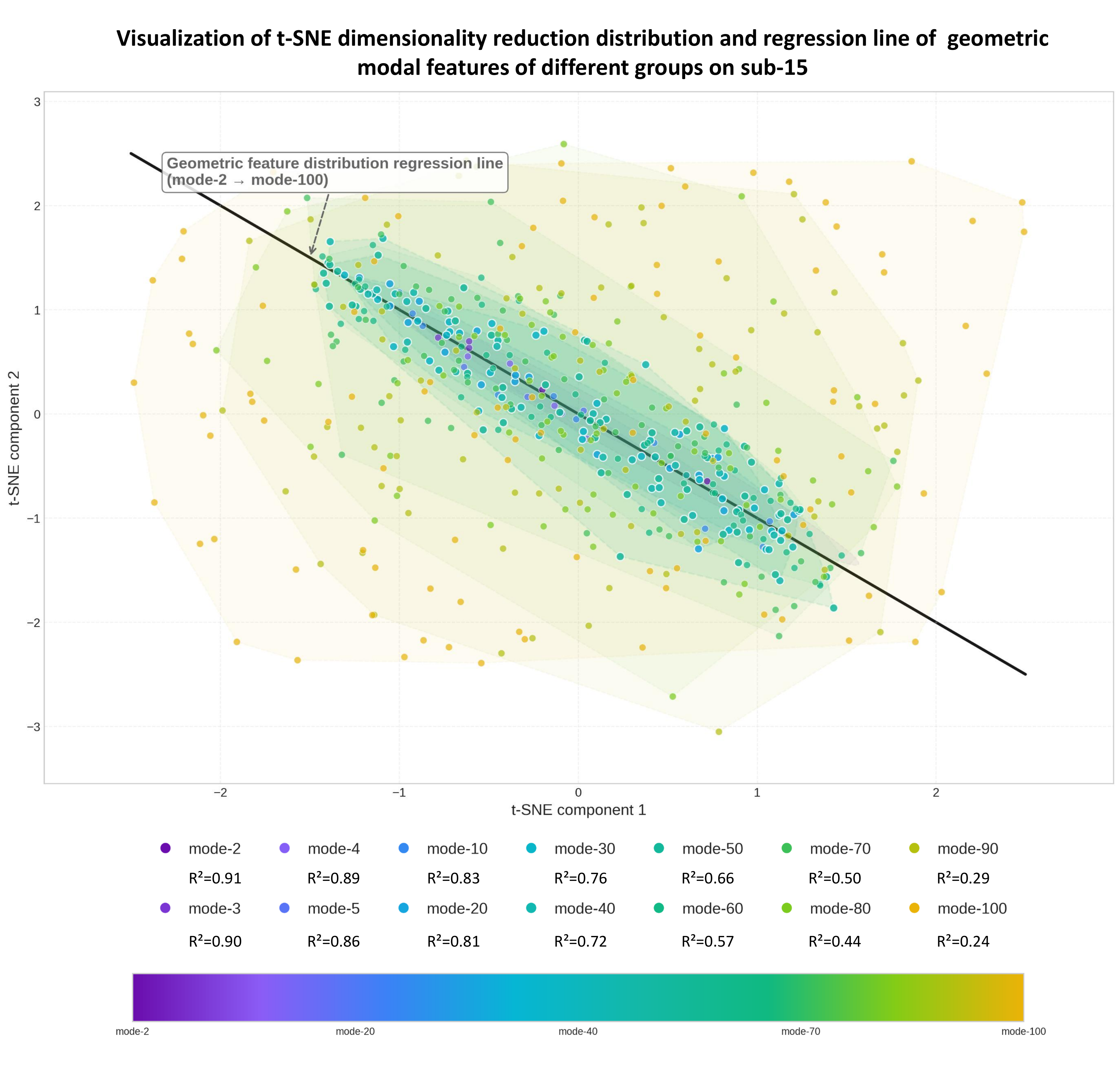}
    \caption{
        Scatter points represent geometric modal feature samples for mode groups (mode-2 to mode-100), color-coded with a purple$\rightarrow$blue$\rightarrow$green$\rightarrow$yellow gradient reflecting the progressive diffusion stage of geometric feature decomposition. The dotted black line represents the principal regression trajectory of the Laplace--Beltrami Operator, serving as a reference axis for quantifying geometric structure deviation. Results show: mode-2 to mode-10 (purple-blue) exhibit highly compact clustering along the diagonal band with strong geometric constraint but limited discriminability; mode-60 to mode-100 (green-yellow) show excessive diffusion beyond the band structure, obscuring geometric-topological relationships; mode-20 to mode-50 (blue-green) maintain band-aligned structure with sufficient inter-sample discriminability while remaining tightly constrained around the regression line, preserving intrinsic cortical manifold geometry. These findings indicate that mode-20 to mode-50 achieve optimal balance between geometric constraint and feature expressiveness, representing the most suitable geometric feature decomposition modes for EEG enhancement tasks.
    }
    \label{fig:sup52}
\end{figure*}

\end{document}